\documentclass[
  journal=pasa,
  manuscript=research-paper, 
  year=202X,
  volume=YY,
]{cup-journal}

\usepackage{amsmath}
\usepackage{amssymb,microtype,siunitx,booktabs}
\usepackage{caption}
\usepackage{subcaption}
\usepackage{graphicx} 
\usepackage{hyperref}
\usepackage{amsmath}
\usepackage{amssymb}
\usepackage{xcolor}
\usepackage{soul}

\usepackage{xcolor}

\definecolor{revisionblue}{named}{cyan}

\newcommand{\rev}[1]{\textcolor{black}{#1}}
 
\title{Constraining the origin of magnetic white dwarfs}

\author{Ananya Mohapatra}
\affiliation{Department of Physics and Astronomy, University of Rochester, Rochester, 14627, NY, USA}
\email[Ananya Mohapatra]{amohapa2@ur.rochester.edu}

\author{Eric G. Blackman}
\affiliation{Department of Physics and Astronomy, University of Rochester, Rochester, 14627, NY, USA}
\alsoaffiliation{Laboratory for Laser Energetics, University of Rochester, Rochester, 14627, NY, USA}

\received {dd Mmm YYYY}
\revised  {dd Mmm YYYY}
\accepted {dd Mmm YYYY}
\published{22 September 202X}


\begin{document}

\begin{abstract}
The origin of magnetic white dwarfs (MWDs) has been a long-standing puzzle. Proposed origin mechanisms have included: fossil fields frozen in from the progenitor convective core; a dynamo in the progenitor envelope; crystallisation dynamos in sufficiently cool white dwarfs; accretion disc dynamos from tidally shredded low-mass stellar or planetary companions, and white dwarf–white dwarf mergers. Here we show how observational constraints on white dwarf magnetic field strengths, ages, and masses can be used to constrain the viability of proposed origin mechanisms. Using data from both an expanded catalogue of 1158 MWDs and a 20 pc volume-limited sample from Gaia DR2, we find that the fossil field mechanism overpredicts the number of magnetic white dwarfs, which suggests that additional constraints beyond just the WD mass being contained in the progenitor convective core are required to determine which WDs retain fossil fields.
Crystallisation dynamos occur too late to explain the bulk of magnetic white dwarfs. With the progenitor envelope dynamos impeded by the theoretical challenge of depositing a field from the envelope to the white dwarf core, the two disc dynamo mechanisms emerge as the field origin mechanisms most resilient to present constraints, with mergers best able to explain the young, high-mass, strongly magnetised MWDs. The methods herein also reveal observational data gaps and motivate future acquisition of more complete data.
\end{abstract}

\section{INTRODUCTION}
\label{sec:int}
Magnetic fields provide an external signature of  internal dynamics that helps to probe  the evolution of stars and compact objects. For white dwarfs (WDs), \rev{we compile a sample of $1158$ magnetic white dwarfs (MWDs) combining the Montreal White Dwarf Database (MWDD) \citep{Dufour+2017}, the hydrogen-atmosphere catalogue of \cite{2023ApJ...944...56A}, and the DESI MWD sample of \cite{2026ApJ..1001..202A}},
with reported magnetic field strengths ranging  from $\sim 10^3$ to $\sim 10^9$ G.

Zeeman splitting of Balmer lines and spectropolarimetry are the primary methods for measuring these WD magnetic fields \citep{Kepler+2013,Berdyugin2022, Berdyugin2023,berdyugin2024searchingmagneticfieldsfeatureless}, which provide a measure of the total magnetic field magnitude averaged over the surface of the WD, $\langle|B|\rangle$.  Spectropolarimetry provides information about the average value of the signed line of sight field on the surface of the WD, $\langle B_z \rangle$, which is a lower limit on the total field. If only the line of sight field is available and dipole topology is assumed, then the total dipole field can be estimated as a lower limit on the total field. Both techniques are more likely to be available for cooler WDs with fields in the range $1 \lesssim\langle|B| \rangle\lesssim 100$MG \citep{Toonen}.

\citet{Hardy_2023} analysed hydrogen-rich MWDs using data from the MWDD supplemented by photometric and spectroscopic data from various sky surveys. 
The MWDD sample focuses on MWDs exhibiting  Zeeman splitting, mostly indicative of magnetic field strengths $>1$ MG. It reveals a notable increase in the number of MWDs above $\sim 0.7 M_\odot$, with the most massive white dwarfs (WDs) exhibiting the strongest field strengths.
Overall, the origin of these magnetic fields is debated \citep{Bagnulo_2022, 2025ApJ...990...25M}. Mechanisms  proposed to account for MWDs can be  classified into five categories: fossil fields frozen in from the progenitor convective core; crystallisation dynamos;  common envelope dynamos; accretion dynamos from shredded companions; or  WD/stellar mergers.
Determining which of these survive evolving 
observational and theoretical constraints 
is important for constraining the physics and evolutionary pathways of WDs. Each of the  mechanisms  raises specific challenges to be addressed.

\cite{2005MNRAS.356..615F} 
 review known properties of  MWDs and draw connections to the properties of their progenitor main sequence stars. They argue that WD magnetic fields may arise from magnetic flux conservation during stellar evolution off the main sequence 
\citep{Ginzburg1964,Woltjer1964,Braithwaite+2004}.  Indeed, main-sequence stars generate magnetic fields through contemporary convective dynamos, and for stellar masses greater than $1.5 M_\odot$ a fossil field could in principle persist in the convective core \citep{Camisassa2024,Castro-Tapia_2026}. 

 Complementarily, WDs undergo crystallisation as they reach sufficiently low temperatures \citep{Kirzhnits1960,VanHorn1968}. As the WD solidifies into a crystalline lattice, heavier elements like $^{16}O$ sink towards the centre, while lighter elements like $^{12}C$ rise towards the outer layers, initiating a composition gradient \citep{Stevenson1980}. The gravitational potential energy released is converted to the kinetic energy of buoyancy, which drives convection. This, along with WD rotation, may sustain a magnetic dynamo to amplify the magnetic field \citep{Ginzburg_2022,Fuentes+2024}.
 Similar core crystallisation has long been thought to  provide an important source of free energy for outer core dynamos in planetary bodies such as Earth \citep{Isern_2017}. But we must assess how many MWDs are old enough to be consistent with the timing of crystallisation onset to be explained this way.

Yet another source of free energy for MWD field amplification is differential rotation that arises when a WD core of a giant star and a companion evolve in a common envelope.  Some of this energy is deposited in the giant envelope, which is a possible dynamo location  \citep{Tout+2008, roepke2022simulationscommonenvelopeevolutionbinary}.
But  the necessary differential rotation free energy  for a dynamo can also arise in a WD merger \citep{10.1093/mnras/stt1910} leading to field amplification not dissimilar from that which likely occurs in neutron star mergers \citep{Kiuchi+2024}.
Alternatively, differential rotation may arise in an accretion disc formed from a tidally shredded companion during or after common envelope evolution \citep{Nordhaus+2006,Nordhaus+2011,Ondratschek+2022}.  These interactions provide the physical conditions where shear and magnetorotational-instability-driven fluctuations can generate large-scale magnetic fields \citep{10.1093/mnras/stw1619}. Even though a companion would play a key role during these amplification processes, it may be gone by the time the MWD is observed. \rev{Mergers also produce isolated WDs \citep{Briggs_2014}}
and most of the stronger MWDs are observed as isolated WDs \citep{Nordhaus+2011}. But do these mechanisms produce sufficiently strong fields to explain all MWDs?

Here we address the above open issues raised  for each of these MWD field generation scenarios and combine  observational data and theoretical constraints to infer the relevance of each mechanism.
\section{RESULTS}
\label{sec:xxx}

\subsection*{Fossil field origin of MWDs } 
\label{sec2}
According to the fossil field hypothesis, the magnetic fields of MWDs observed today would be 
flux-frozen remnants from the progenitor
main-sequence star. This  presumes that magnetic flux from a core dynamo can be retained as the  WD is formed.
Convective core dynamos generate magnetic fields in main-sequence stars with masses $M/M_\odot>1.2$. For stars with  $0.3<M/M_\odot<1.2$,  convection occurs only in the envelope.
Stars with  $M/M_\odot <0.3 $ are fully convective \citep{2013sse..book.....K, 1997A&A...327.1039C}. 

The  fossil field mechanism requires at least three conditions to be met. First, the WD mass must not exceed the convective core mass of the progenitor to ensure the magnetic field is generated throughout the region that becomes the WD. Second, the field must survive the formation phase of the WD. This is threatened if the conditions for dynamo growth become disfavoured before the convective turbulence fully shuts down.  Third, the field produced must extend to the WD surface.

\begin{figure}[H]
  \centering
  \begin{subfigure}{1\textwidth}
 \caption{}
 \includegraphics[width=\linewidth]{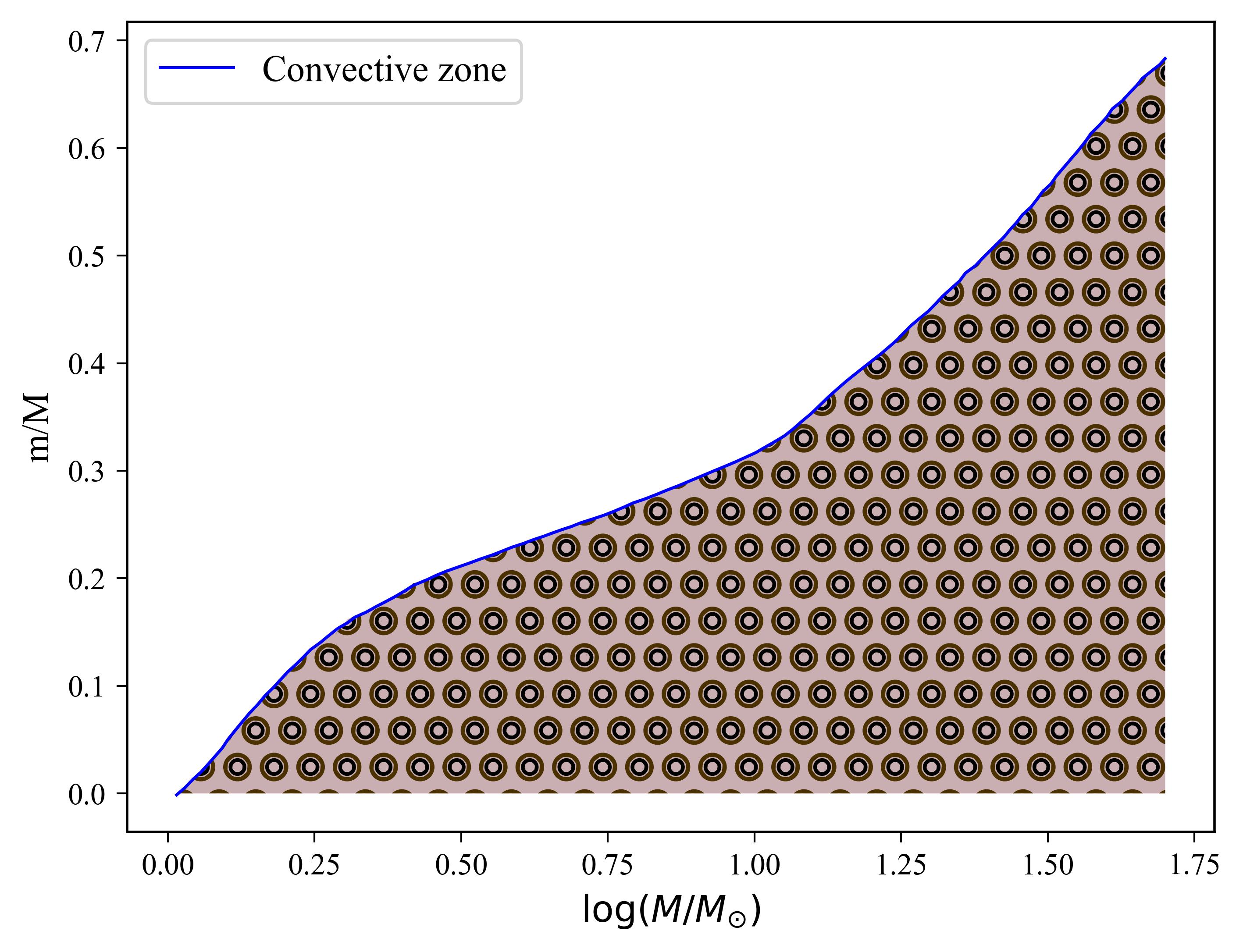}
  \label{fig1a}
  \end{subfigure}

  \begin{subfigure}{1\textwidth}
  \caption{}
\includegraphics[width=\linewidth]{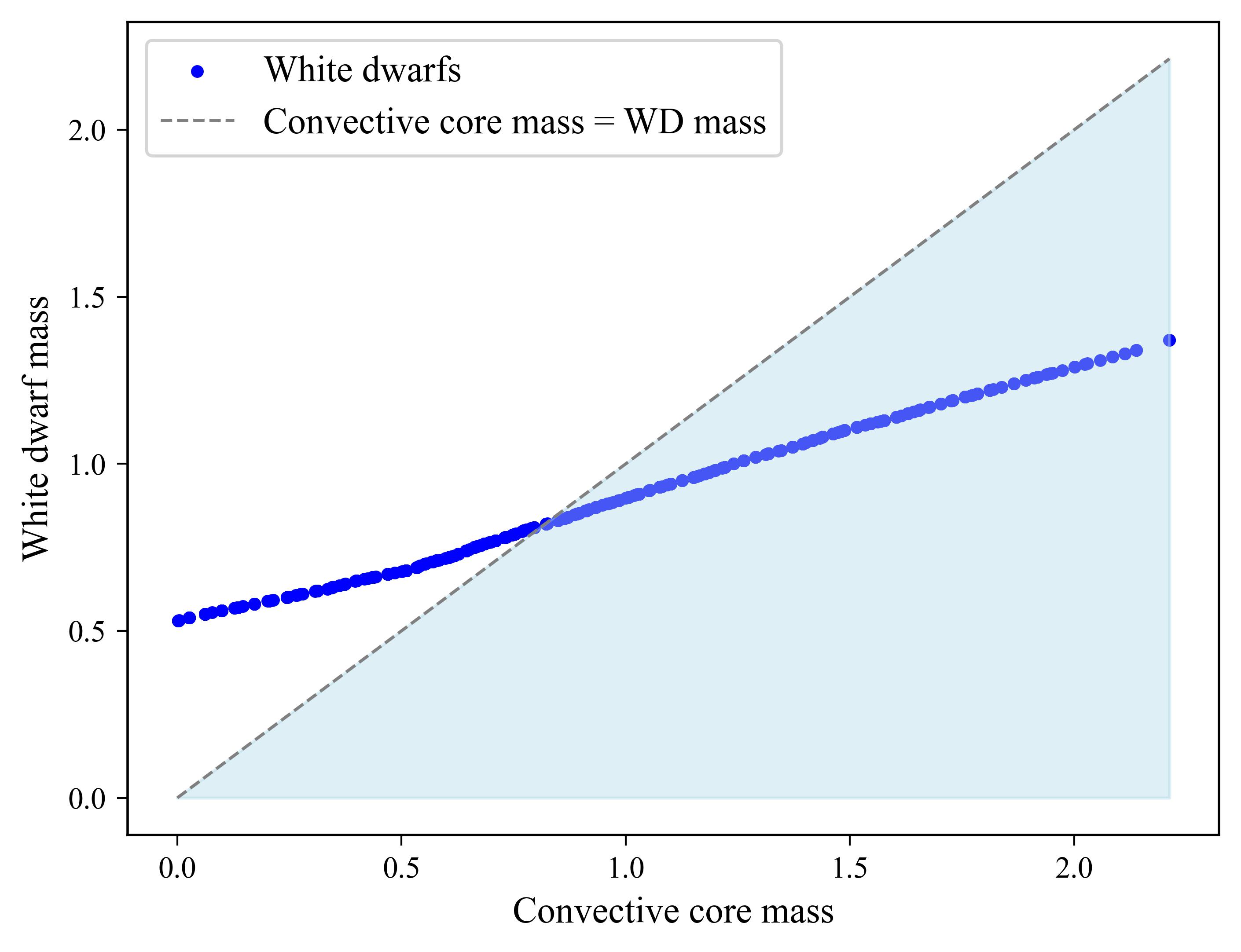}
  \label{fig1b}
  \end{subfigure}
  \caption{\textbf{Comparison of WD Masses with Convective Core Mass} \textbf{(a)} Fraction of a main sequence star's mass within its convective core `m/M' versus Logarithm of total mass `$\log(M/M_{\odot})$'. The shaded region represents the convective zone. Adapted from Figure 22.7 of Kippenhahn and Weigert.
\textbf{(b)} Comparison of WD masses to  progenitor convective core masses (blue dots). Those WDs to the right of the `Convective core mass = WD mass' line fall entirely within the convective zones of their progenitors.}

\end{figure}   

To address the  first condition,  we estimate the mass of the main sequence star from which the WD evolved. Based on the semi-empirical initial-final mass relationship described by \citet{Catalan+2008}, the 
 potential progenitor star's mass $\left(M_i\right)$ for each WD is calculated differently according to the WD mass categories. For WDs with masses $M_\text{WD}\leq 0.6882 M_\odot$, the relationship is $M_\text{WD} = 0.096 M_i + 0.429$. For WDs with masses $M_\text{WD} > 0.6882 M_\odot$, the relationship is $M_\text{WD} = 0.137 M_i + 0.318$. These linear relations are derived from a weighted least squares linear fit of observational data, providing a more accurate model for the initial-final mass relationship in different WD mass ranges. 
Figure \ref{fig1a} shows the mass fraction of the convective region relative to the total progenitor stellar mass \citep{KippenahanWiegert}.
We  use this to compare the initial progenitor mass with the final WD mass in Figure \ref{fig1b} and show which WDs fall within their progenitor's convective core mass region. The WD data are represented by blue dots, and the dashed line delineates the boundary at which WD mass equals that of its progenitor's convective core. WDs below this line meet the condition of forming from mass entirely contained within the convective core.

We next address the aforementioned second condition for a viable fossil field,  that the WD must form before the  magnetic field decays. Large scale convective or turbulent dynamos require not only macroscopic transport  but also a sufficiently high magnetic Reynolds number for the large-scale field stretching to overcome microphysical dissipation \citep{ChrisScaling}. The magnetic Reynolds number measures the relative importance of the advective transport of the magnetic field by the flow compared to resistive diffusion at the onset of convection and is expressed as,
\begin{equation}
R_m=\frac{v_{conv} \ell}{\eta},
\end{equation}
where $v_{conv}$ is the characteristic convective velocity, $\ell$ is a characteristic convective length scale and $\eta$ is the microphysical magnetic diffusivity.
Turbulence involves a growth transport term, such as  kinetic helicity, to overcome turbulent diffusion \citep[e.g. for a review see][]{Brandenburg_2005} when the dynamo sustains.  \rev{If the dynamo shuts down just when convection shuts down,
then the field would persist, as it would decay very slowly through microphysical resistive diffusion.  If the dynamo shuts down because the growth transport term is suppressed before turbulent diffusion is fully suppressed, then 
the system would still retain some turbulent diffusion that could, in principle, erase the magnetic fields before the white dwarf fully forms.
We estimate, however, that by the time the convective velocities have dropped to sufficiently small values for the dynamo to shut off, the resulting turbulent diffusion would be too weak to significantly damp the field before convection stops entirely.}
In short,  a significant fossil field could seemingly survive.

\rev{An observable fossil field  therefore comes down to the third aforementioned condition: }
whether the magnetised core extends to the surface  of the  emergent WD or remains buried  within. \rev{This constraint, as observed in Figure ~\ref{fig1b}, favours the detection of fossil-fields only in MWDs above $M\simeq0.82M_{\odot}$, rather than the full MWD population.} Since MWDs formed through such mechanisms would appear magnetised at early cooling ages, objects consistent with a fossil field origin would be young, massive MWDs.
 As shown by \citet{Camisassa2024}, the detailed evolution of the convective boundary is likely complex, with the core convection zone shrinking during the main sequence phase, and the outer convective envelope later penetrating inward towards the core.  \rev{However, 
 fields anchored in regions that do not ultimately become part of the WD interior are less relevant to its subsequent magnetism.}

 Recent asteroseismic studies suggest strong internal magnetic fields in the $\approx 0.2 M_\odot$ helium cores of red-giant stars, the material that later becomes the white-dwarf interior \citep{SeismoGangLi, Fuller_2015}. Although these detections confirm that strong magnetism can exist before envelope loss, evolutionary calculations show that such fields are likely to remain buried beneath the surface layers and therefore undetectable at the surface in young WDs  \citep{Camisassa2024}. \rev{The spatial extent of the internal field is therefore important. \cite{2026A&A...708L..14E} distinguish between fields generated solely during main-sequence core convection and internal fields that fill the radiative interior as the star evolves towards the WD stage. Since the magnetic fields are no longer limited by the size of the convective core in the latter case, the burial problem would be reduced.  A fossil field extending through the radiative interior would thus be required to produce observable MWDs with masses below $0.82M_\odot$. } 
 
In addition to magnetic fields generated by convective core dynamos during the main sequence, some upper main sequence  Ap/Bp stars,  for example,
exhibit strong, highly structured fields \citep{2008A&A...481..465L} already present at the zero-age main sequence (ZAMS). These are thought to be  inherited from earlier pre-main sequence stages of star formation and may persist
in stable, non-convective envelopes. \rev{Where these fields are anchored may determine their subsequent influence on WD fields. 
If these fields are anchored in the envelope, then they would be detached from the convective cores that become the WD and would be lost when the envelope is lost. If they are instead anchored within the  convective stellar core that contracts to form the WD, such fields could in principle survive  WD formation, although they would likely be dominated by the fields generated   in situ from core convection.}

If residing below the line in Figure \ref{fig1b} were the only condition for a WD to retain a fossil field generated in the core, that would over-predict the number of magnetised WDs found in surveys for high-mass WDs. In our sample, $381$ MWDs meet this condition; however, close to $5000$ WDs from the total WD survey have their masses contained within the convective core region of their progenitors. 
\rev{A viable fossil-core field paradigm must therefore involve  multiple filters before the WD is observed as magnetic. As alluded to above, the first is that the field must avoid being erased by turbulence during the course of WD formation, that is, there can be no  significant phase where turbulent diffusion wins over dynamo growth. 
The second filter is overcoming field burial: even if the containment condition as shown in Figure \ref{fig1b} is met and  the WD is formed entirely from the magnetised convective core, this does not guarantee that the field immediately threads the observable surface. 
The WD could harbour an internal fossil field that only appears after some relaxation or cooling time. The final filter is detectability: even if a surface field is present, the magnetic-field detection biases due to mass, temperature, field strength, and atmospheric properties could make some fossil fields difficult to measure.}



\subsection*{Crystallisation dynamos may explain some but not all MWDs}
\label{sec3}
The exact temperature evolution of WD crystallisation depends on the thermodynamic properties and composition of the WD, including the  carbon-to-oxygen ratio. For typical WDs, crystallisation is expected to take place within a temperature range of  $10^4 \lesssim T\lesssim 10^5$K \citep{Schreiber}.
We can rule out crystallisation dynamos as the primary explanation for MWDs younger than the age of crystallisation, and check for evidence of an increase in MWDs after crystallisation \citep{Blatman2024}. To determine crystallisation ages, we use the theoretical evolutionary sequences of \citet{2020ApJ...901...93B}. 
These sequences\footnote{accessible online at \href{http://www.astro.umontreal.ca/~bergeron/CoolingModels/}{http://www.astro.umontreal.ca/bergeron/CoolingModels/} under the section titled `Evolutionary Sequences'} 
translate the effective WD temperature into a cooling age. We use them  to construct temperature-age cooling curves \citep{ginzburg2024youngerageoldestmagnetic} for different WD masses, determine the corresponding ages of crystallisation  \citep{Tremblay_2019, 2025ApJ...990...25M}, and compare these to the observed 
MWDs. 

\begin{figure*}[t!]
    \centering
    \begin{subfigure}[t]{0.90\textwidth}
        \centering
        \includegraphics[width=\linewidth]{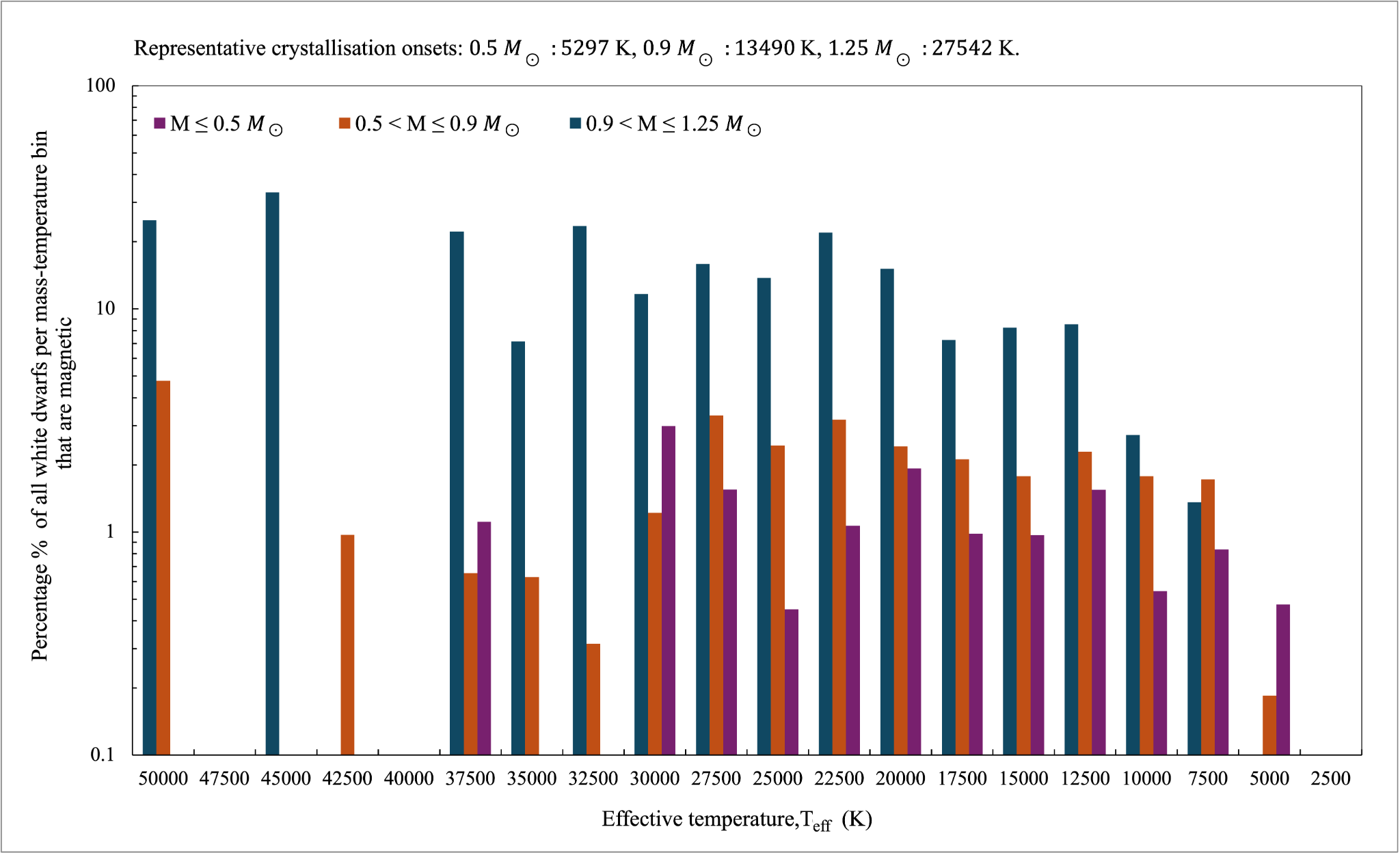}
        \caption{}
        \label{fig:WDpercent_top}
    \end{subfigure}

    \vspace{0.5cm}

    \begin{subfigure}[t]{0.51\textwidth}
        \centering
        \includegraphics[width=\linewidth]{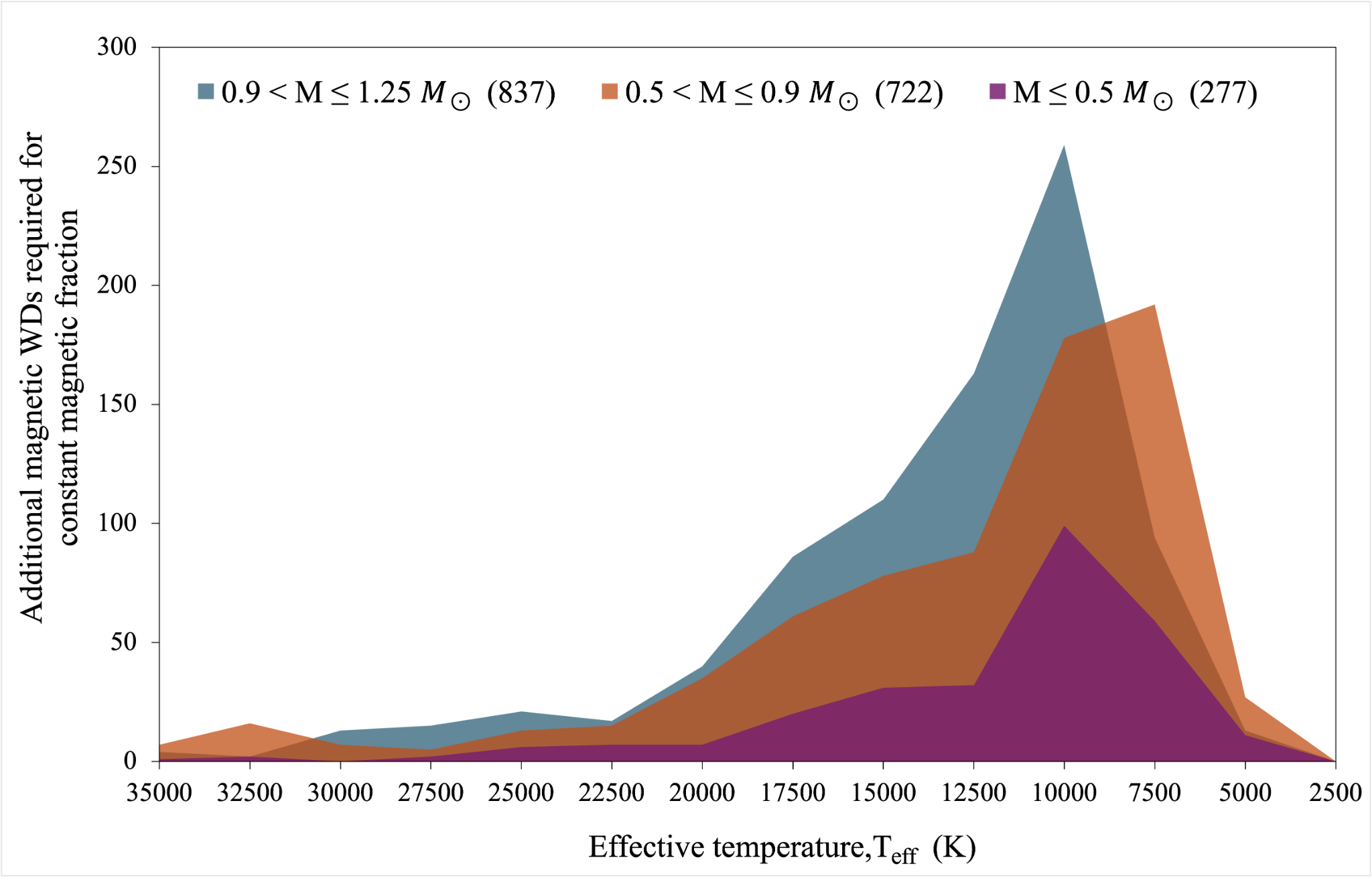}
        \caption{}
        \label{fig:WDpercent_bl}
    \end{subfigure}
    \hfill
    \begin{subfigure}[t]{0.48\textwidth}
        \centering
        \includegraphics[width=\linewidth]{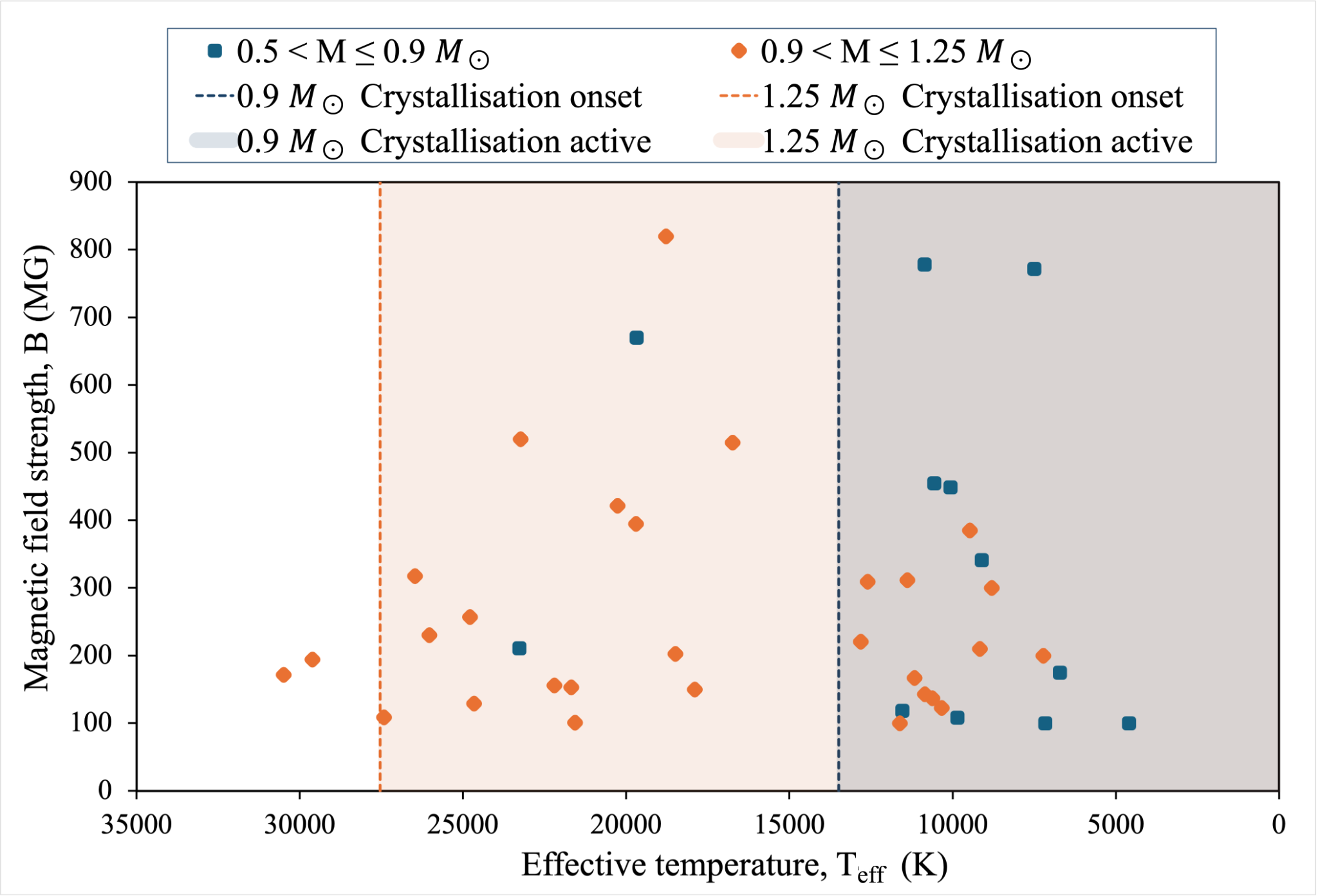}
        \caption{}
        \label{fig:WDpercent_br}
    \end{subfigure}

    \caption{\textbf{Incidence of magnetic white dwarfs (MWDs) across mass and temperature.}
    \textbf{(a)} Fraction of WDs in each mass--temperature bin that are magnetic, with bars colour-coded by mass range. The bins run from hotter, younger WDs on the left to cooler, older WDs on the right. 
    \textbf{(b)} Estimated number of additional MWDs required in each temperature bin to maintain a constant magnetic fraction along the cooling sequence for each mass range. 
    \textbf{(c)} High-field MWDs ($>100$ MG) shown as a function of effective temperature. The dashed vertical lines mark the crystallisation-onset temperatures, and the shaded areas indicate the corresponding crystallisation-active regions for selected WD masses.  
    }
    
    \label{fig:WDpercent}
\end{figure*}

Figure \ref{fig:WDpercent_top} shows the percentage of WDs that are magnetic in each mass–temperature bin, with three different colours representing three distinct mass ranges. This percentage for each bin is estimated as:
\begin{equation}
\operatorname{MWD}_{\left(M_i,T_j\right)}=\frac{N_{\mathrm{MWD}}\left(M_i, T_j\right)}{N_{\mathrm{WD}}\left(M_i, T_j\right)} \times 100 \%
\end{equation}
$N_{\mathrm{WD}}\left(M_i, T_j\right)$ is the total number of white dwarfs (magnetic + nonmagnetic) in that bin, and $N_{\text {MWD }}\left(M_i, T_j\right)$ is the number of magnetic white dwarfs in the same bin. Here, $i$ cycles through the three mass ranges, each shown by a different bar colour, while $j$ steps through the temperature bins along the x axis. To pinpoint any ($M_i, T_j$) bin, simply match the colour of the bars to fix $i$ and then read its horizontal position to set $j$. The temperature axis traces a cooling sequence, from hotter WDs on the left to cooler WDs on the right.


\rev{If magnetic fields in MWDs do not decay appreciably over the entire temperature range shown in Figure \ref{fig:WDpercent_top}, 
then any additional mechanism that turns on during WD cooling would increase the 
number of MWDs towards cooler temperatures because cooler WDs were once hotter WDs.
Figure \ref{fig:WDpercent_bl} shows the additional number of MWDs required in each temperature bin to maintain a constant magnetic fraction for each mass range. The higher-mass range, shown in blue, generally requires the highest deficit correction, compared to the lower-mass ranges, shown in orange and purple. This suggests that many MWDs are missing from what would be a complete observational sample.} 

\rev{To assess whether crystallisation could play a role in high-field MWDs, we compare MWDs with $B> 100 \text{ MG}$ to the crystallisation-active regions for representative WD masses. Figure \ref{fig:WDpercent_br} shows that several of these MWDs lie within or near the crystallisation-active region, a necessary condition for crystallisation to potentially contribute to their magnetic evolution.}


\begin{figure*}[t!]
    \centering
    \begin{subfigure}[t]{0.48\textwidth}
        \centering
        \includegraphics[width=\linewidth]{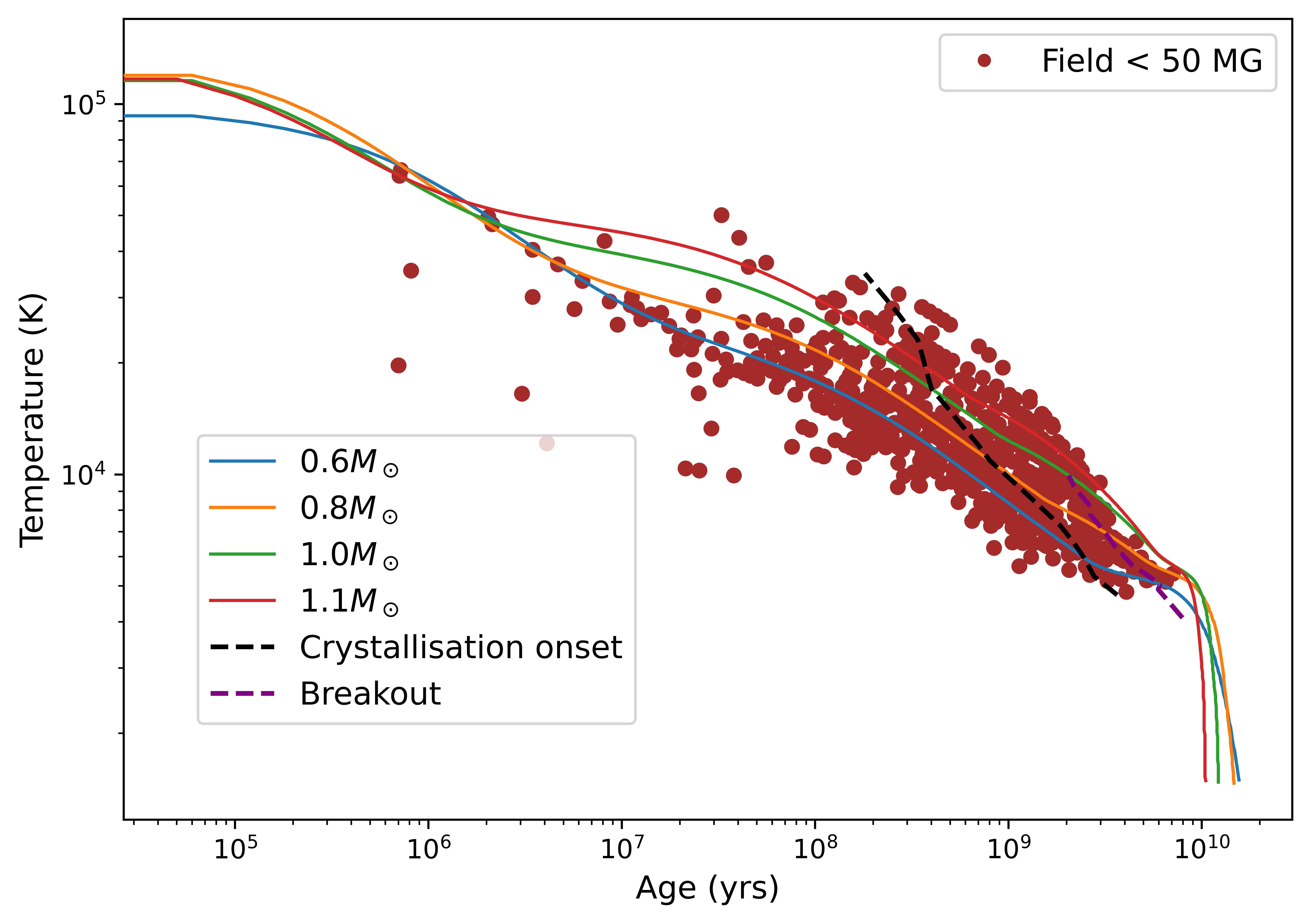}
        \caption{}
        \label{field<50}
    \end{subfigure}
    \hfill
    \begin{subfigure}[t]{0.48\textwidth}
        \centering
        \includegraphics[width=\linewidth]{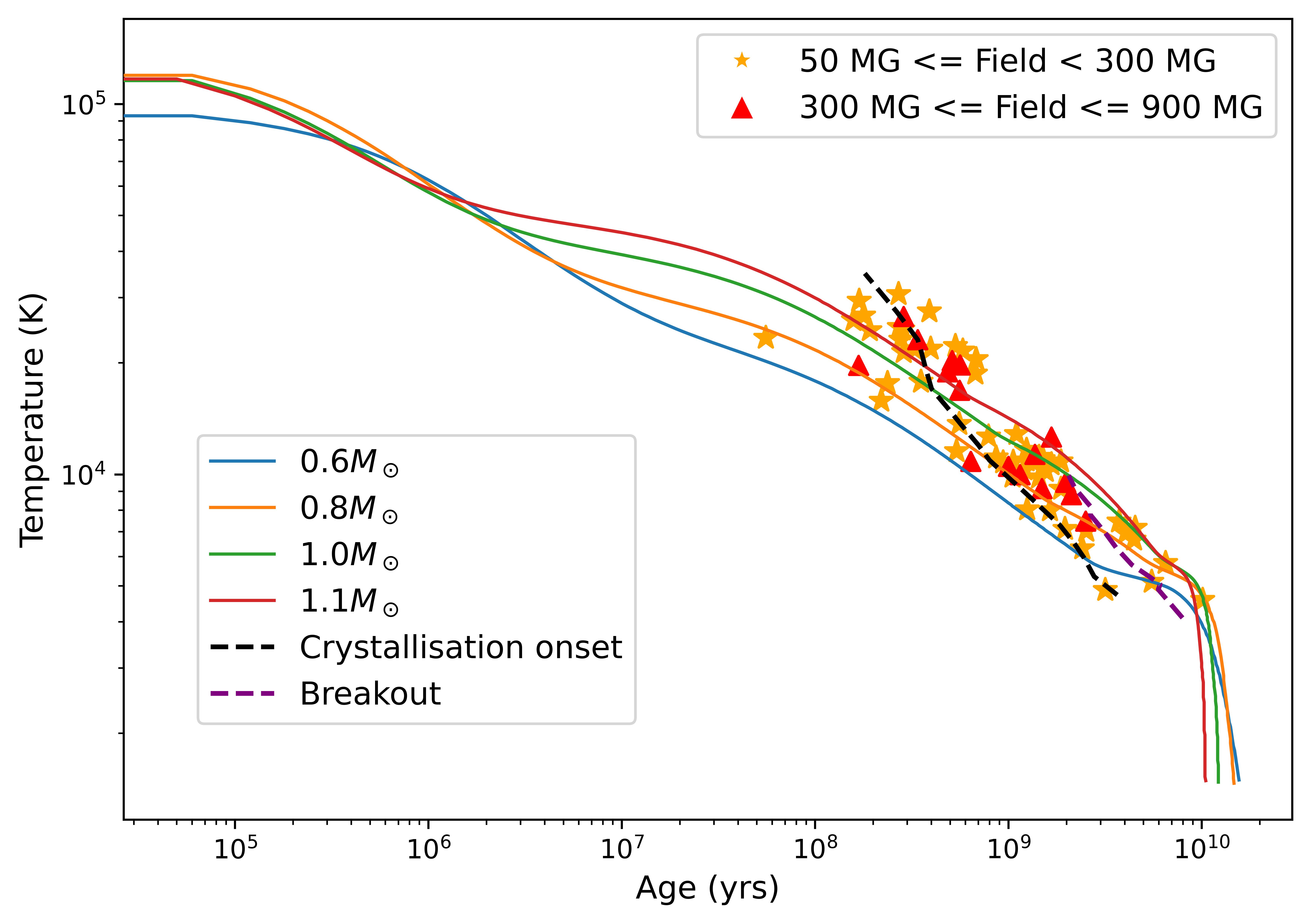}
        \caption{}
        \label{field50-900}
    \end{subfigure}
    \caption{\textbf{Cooling curves for representative WD masses, with the onset of crystallisation and magnetic field breakout times marked by black and purple dashed lines, respectively.} \textbf{(a)} MWDs with detected field strengths below 50 MG. \textbf{(b)} MWDs with field strengths between 50–300 MG and 300–900 MG. Intersections with the dashed curves indicate the corresponding evolutionary stages for each mass.}
    \label{fig:CC_MWDs50-900}
\end{figure*}

\begin{figure}[t]
    \centering
    \includegraphics[width=1\columnwidth]{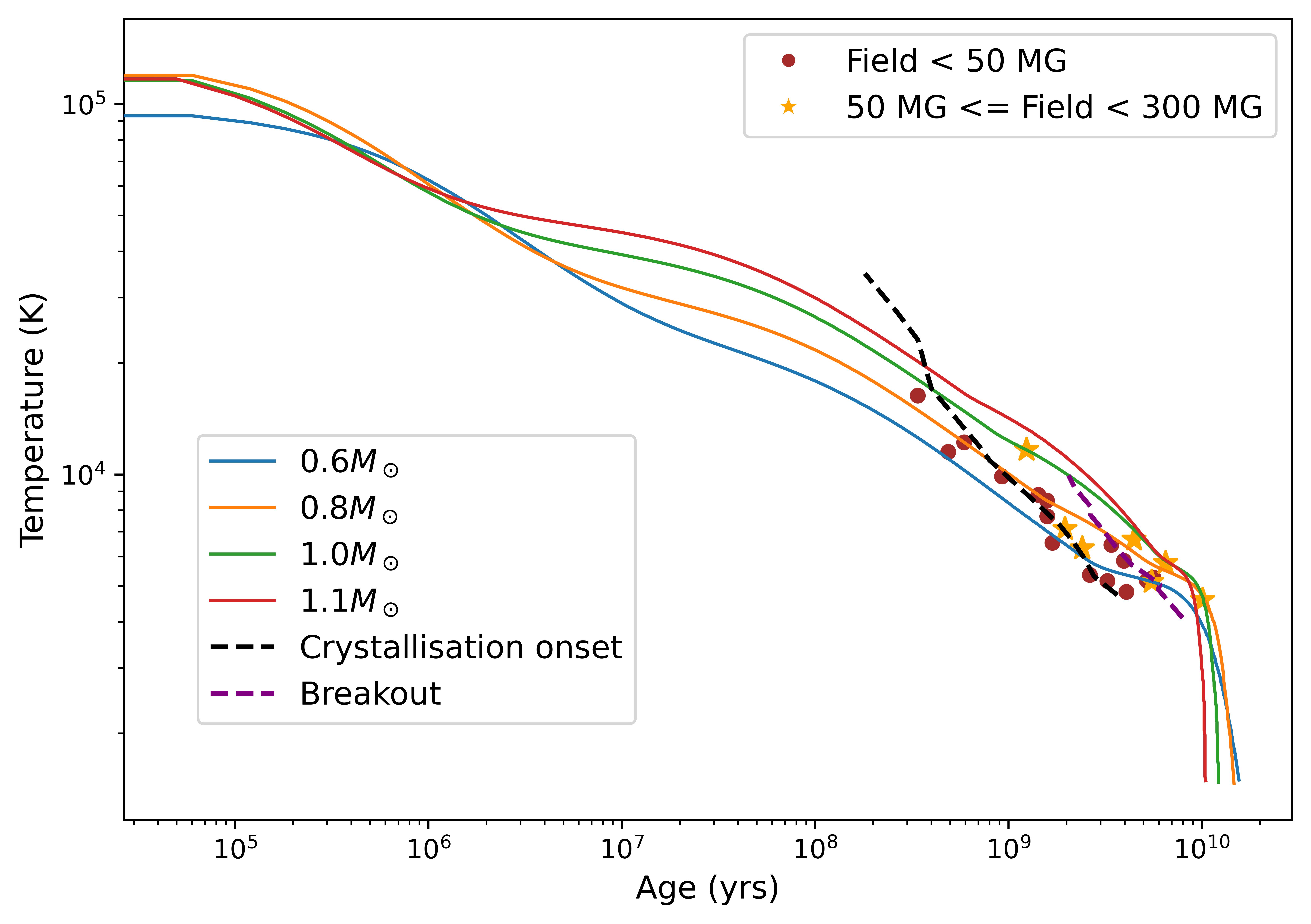}
    \caption{\textbf{Cooling curves for representative WD masses, with onset of crystallisation and magnetic-field breakout times marked in the volume-limited sample.} MWDs with field strengths below 50 MG and in the range 50--300 MG are shown relative to these stages. }
    \label{fig:CC_MWDs50-900_VL}
\end{figure}



To complement the temperature-based comparison in Figure \ref{fig:WDpercent}, Figure \ref{fig:CC_MWDs50-900} places the MWDs on cooling-age curves for representative WD masses, \rev{and marks both the onset of crystallisation and magnetic-field breakout times. The marker for  crystallisation onset indicates when a crystallisation dynamo could begin, and the breakout time indicates when a magnetic field generated by such a dynamo could reach the observable surface.}
In Figure \ref{field<50}, we show MWDs with field strengths below $50 \text{ MG}$ along these curves,  relative to these markers. Figure \ref{field50-900} presents similar cooling curves for two higher field strength ranges, $50-300 \text{ MG}$ and $300-900 \text{ MG}$, marked with different symbols. 
MWDs older than the age of crystallisation onset satisfy the necessary condition for a crystallisation dynamo to have influenced their magnetic-field evolution, while objects lying beyond the breakout curve are more directly consistent with an observable crystallisation-generated surface field. 
Overall, the data and figures reveal a modest bump in the fraction of magnetised WDs near or after the onset of crystallisation, suggesting a potential link to the crystallisation dynamo. This is further supported by Figure \ref{fig:WDpercent_br} and the cooling curves in Figures \ref{field50-900} and \ref{fig:CC_MWDs50-900_VL}, which show that several high-field MWDs appear within or near the crystallisation-active regions.  
\rev{This association  suggests that crystallisation cannot   be ruled out as an  influence on magnetic-field evolution for current data \citep{2023ApJ...944...56A}. However, the interpretation  is  complicated due to the fact that crystallisation causes cooling delays, increasing the time WDs spend in a certain temperature range. This allows  pre-buried fields more time to become detectable, which could also contribute to the observed bump without a crystallisation dynamo.} 

To address potential biases in the full MWD sample \citep{Bagnulo_2022}, we repeat our analysis using the volume-limited sample of 20 pc from Gaia DR2 \citep{10.1093/mnras/sty2057}. This subset provides a more complete local representation of WDs across a range of ages. We compare the results from this volume-limited sample, shown in Figure \ref{fig:CC_MWDs50-900_VL}, with those from the full MWD sample to ensure that any data limitations affecting our conclusions are properly accounted for. \rev{In the full sample, high-field MWDs ($>50 \text{ MG}$) are largely concentrated at cooler temperatures, and many lie within the crystallisation-active regions. In contrast, the low-field ($<50 \text{ MG}$) population includes a stronger representation of hotter, younger objects well before crystallisation occurs. The 20 pc sample shows a similar but more limited comparison: most of the MWDs shown lie within or near the crystallisation-active regions across the presented field strength ranges. Thus, the volume-limited sample is broadly consistent with the full-sample trend that crystallisation may be relevant for some cooler high-field MWDs.} 
However, we reiterate that the presence of hot magnetised WDs in the full sample before the onset of crystallisation demonstrates that this cannot be the only mechanism responsible for the magnetic fields of all MWDs. 

\rev{The inclusion of field breakout times in Figure \ref{fig:CC_MWDs50-900_VL} tightens the timing constraint on crystallisation dynamos. Several MWDs lie beyond the onset of crystallisation, but fewer objects satisfy the stronger condition of being older than the breakout time needed for the field to become observable.}

Importantly,  there remains the ongoing challenge of determining what field strengths  crystallisation dynamos actually produce.  Some theoretical models \citep{Blatman2024} and simulations \citep{Castro_Tapia_2024} of the crystallisation dynamo predict magnetic field strengths that are too weak   to account for most observed MWDs, while \cite{Fuentes+2024} argue that it may be possible. \rev{Since the magnetic-field strength predicted for a crystallisation dynamo depends strongly on the convective velocity and mixing length during phase separation, \cite{2024ApJ...961..197M} re-examined the assumptions made in earlier cooling models. They argued that the uniform-mixing prescription presumes large-scale turbulent motions in a way that is not self-consistent, and found instead that the relevant velocity-length product is many orders of magnitude smaller when neutrally buoyant or thermohaline mixing is considered. This leads to predicted fields of order $\lesssim 10 \text{ kG}$, far below the MG fields observed in most MWDs.}

\subsection*{The trouble with envelope dynamos}
\label{sec4}

Dynamos in giant stellar envelopes are another possible source of free energy to generate magnetic fields that could in principle be used for the magnetic fields of WDs.  In particular,  large-scale magnetic field amplification in the envelope of the giant star during a common envelope (CE) phase has been studied \citep{Regos+1995,Nordhaus+2007,Tout+2008,Kissin+2015}. The CE phase typically occurs when a red giant or asymptotic giant branch star engulfs a companion. As the companion star inspirals, it transfers orbital energy and angular momentum to the envelope, causing differential rotation and thus providing a source of free energy in  shear. Because the envelopes of giant stars are strongly convective, conditions for an $\alpha-\Omega$  type dynamo are favourable, with the highest amplification  at the interface between the convective and radiative regions, where the shear is strongest.

However, as the CE phase progresses and the envelope is drained, the mechanism faces challenges \citep{Nordhaus+2011}.
To reach the WD surface, the magnetic fields must diffuse or be pumped downward through the layers of the evolving star while the envelope is expanding and escaping outward.
Moreover, the envelope dynamo operates only transiently during the CE phase.  For the magnetic fields to be inherited by the WD, they must be available for deposit until the WD fully forms.
Unless these challenges can be resolved convincingly, these models are unlikely to be the dominant source of MWDs.

\section*{Magnetic field from accretion of a shredded companion}
\label{sec5}



An accretion disc around a WD can  form from the CE interaction if a low-mass companion star or planet is tidally disrupted by the WD core of the  primary giant star \citep{Nordhaus+2006,Guidarelli+19}.  WD fields can result from accreting the magnetised material \citep{Nordhaus+2011}.


To estimate the total and poloidal field strengths that result from accretion and compare them to values observed in the  MWD population,  we consider companion masses ($M_\text{c}$) ranging from sub-Jupiter-mass ($<M_\text{J}$) planets to low-mass stars and first estimate the fields generated by a dynamo operating within the associated disc. 
An accretion dynamo can operate within the disc of shredded companion material, which can survive evaporation within the envelope  \citep{Guidarelli+19} to accrete flux.
For example, MRI turbulence  \citep{Salmeron2007MNRAS.375..177S}
can amplify the magnetic field and source the transport coefficients for a large scale $\alpha-\Omega$-type dynamo in a stratified disc.

We assume a  standard  accretion disc model \citep{Shakura1973, Frank+2002} for  which the mean
fields and flows are axisymmetric, with turbulent transport supplied by the MRI \citep[e.g.][]{Balbus1998}.
 The Alfvén speed at radius $R$  for the total magnetic field in an MRI-saturated disc can be estimated from $\alpha_{ss}\beta
 \sim 0.1$ \citep{Blackman+2008}, where $\beta$ is the ratio of thermal to magnetic pressure, and $\alpha_{ss}$ is the Shakura-Sunyaev dimensionless parameter quantifying the turbulent stress that transports angular momentum. 
 Typically,   $0.0001 \le \alpha_\text{ss} \le 0.1 $ 
 as measured in  simulations or inferred from observations.
 Since the magnitude of the mean field dynamo-generated mean magnetic field in the disc is less than the total field  magnitude,  we have a relationship $\alpha_{ss}{\beta({\overline v}_A}) \geq 0.1$ which implies that the squared Alfvén speed for the mean field is approximately ${\overline v_A}^2 \leq 20 \alpha_{ss} c_\text{s}^2$  \citep{Blackman_2001}, where the overline denotes quantities associated with the mean magnetic field.

 Here, $c_\text{s}$ is the sound speed which, from hydrostatic equilibrium, can also be expressed as the product $c_\text{s} \simeq H \Omega_\text{k}$ where $H$ is the disc scale height, $\Omega_\text{k}$ is the Keplerian orbital frequency. From local mass conservation, the mass density of the disc in terms of the local accretion rate $\dot M$  is 
 \begin{equation}
 \rho =\frac{\dot{M}}{4 \pi H R \overline{v}_R}= \frac{\dot{M}}{4 \pi H^3 \alpha_\text{ss} \Omega_\text{k}}
 ,\label{2a}
 \end{equation}
 where $\overline{v}_R\simeq \alpha_{ss}c_s H/R$ is the radial accretion speed. We will make use of equation (\ref{2a}) exclusively at the inner edge of the disc. 

 We use the total magnetic field $B_d$ as an upper limit for the mean toroidal field $\overline{B}_{\phi}$, which dominates the mean.
 Combining the  
 $\alpha_{ss}\beta \sim 0.1$ relation with equation (\ref{2a}) we then obtain
\begin{equation}
    \overline{B}_\phi \leq B_d\simeq \left(80\pi \rho c_\text{s}^2 \alpha_\text{ss}\right)^{1/2} = 4.47 \left(\frac{\dot{M} \Omega_\text{k}}{H}\right)^{1/2}.
    \label{toroidal_field}
\end{equation}

The mean poloidal field, $\overline{B}_p$ is related to the mean toroidal field $\overline{B}_\phi$ according to $\overline{B}_p = \alpha_{ss}^{1/2}\overline{B}_\phi$ \citep{Blackman_2001}. We can  estimate this at the innermost radius of a WD accretion disc,  assuming that the inner edge of the disc extends close to the WD radius.   Scaled for a WD of mass $M_{\text{WD}}$ and radius $R_{\text{WD}}$, this gives 

\begin{equation}
\begin{aligned}
\overline{B}_\text{p} \leq\ & 4.47\alpha_{\mathrm{ss}}^{1/2}\left(\frac{\Dot{M}\,\Omega_\text{k}}{H} \right)^{1/2} \\
\leq\ & 22.62\ \text{MG} \left(\frac{\eta_{\mathrm{ac}}}{0.1}\right)^{1/2}
\left(\frac{\alpha_{\mathrm{ss}}}{10^{-2}}\right)
\left(\frac{M_{\text{c}}}{30 M_{\text{J}}}\right)^{3/4} \\
& \times \left(\frac{M_{\mathrm{WD}}}{0.6 M_{\odot}}\right)^{1/4}
\left(\frac{r_{\text{c}}}{r_{\text{J}}}\right)^{-1/4}
\left(\frac{H/R}{0.05}\right)^{1/2}
\left(\frac{R}{10^{9} \mathrm{~cm}}\right)^{-3/4},
\end{aligned}
\label{eq:Polfield}
\end{equation}
where we have used the accretion rate from viscous angular momentum transport  given by
\citep{Nordhaus+2011}
\begin{equation}
    \Dot{M} \sim 0.07 M_\odot {\rm yr}^{-1} \left(\frac{\Tilde{\eta}}{0.1}\right)\left(\frac{\alpha_{\mathrm{ss}}}{10^{-2}}\right)\left(\frac{H / R}{0.05}\right)^{2}, \label{eq:accrate}
\end{equation}
and $  \Tilde{\eta} = \eta_\text{ac} \left(\frac{M_{\text{c}}}{30 M_{\text{J}}}\right)^{-3 / 2} \left(\frac{r_{\text{c}}}{r_{\text{J}}}\right)^{1 / 2},$
where
$r_\text{J}$ is a Jupiter  radius.
Here $\eta_{\text{ac}}$ is an efficiency parameter that is smaller when more mass is lost to outflows.
Our use of $\Tilde{\eta}$ allows us to control the accretion rate in a form that is independent of 
companion mass and radius in what follows. The aspect ratio $H/R$ in equation (\ref{eq:accrate}) denotes the scale height of the accretion disc relative to its radius and is calculated from hydrostatic equilibrium 
 as 
\begin{equation}
\begin{aligned}
\frac{H}{R} &= \frac{\sqrt{kT/\mu m_\text{p}}}{\sqrt{GM_\text{WD}/R_\text{WD}}} \\
&= \left(\frac{k}{\mu m_\text{p}}\right)^{1/2}  
\left(\frac{3}{\sigma} \cdot \frac{\Dot{M} R_\text{WD}}{G^3 M_\text{WD}^3} \right)^{1/8} \\
&\approx 0.0367 \left(\frac{M_\text{WD}}{0.6 M_\odot}\right)^{-3/8}
\left(\frac{\Dot{M}}{\Dot{M}_\text{ed}} \right)^{1/8}
\left(\frac{R_\text{WD}}{10^9\ \text{cm}}\right)^{1/8}.
\end{aligned}
\end{equation}
Here, $T$ is the disc mid-plane temperature which can be expressed as a fraction $f$ of the virial temperature, $T=
f\,\frac{G M_\mathrm{WD}\,\mu m_p}{k\,R_\mathrm{WD}}$, where $f\le 1$ is a constant, $m_\text{p}$	is the proton mass, $k$ is the Boltzmann constant, $\mu$ is the mean molecular weight, and   $G$ is the gravitational constant \citep{Frank+2002}.
Simulations by \citet{Zhu20132013ApJ...767..164Z} show that the mid-plane gas quickly reaches $f\approx 0.2$ of the maximum virial temperature in WD merger discs, which is likely larger than that in discs formed from smaller companions.  

Accretion of a shredded companion and magnetic field onto a WD at the initial accretion rate could continue for about a viscous time scale, unless  the accumulated stress from the total WD magnetic field, $B_\text{WD}$  counterbalances the Keplerian ram stress exerted by the orbiting material in the  disc first. As we now show, this balance does indeed occur before a viscous time, and therefore provides the limiting field.
Using pressure balance as a rough proxy for stress balance \citep{Blackman_2001}, the magnetic pressure
$\frac{B_{\text{WD}}^2}{8 \pi}$ balances the Keplerian ram pressure $\rho V_k^2$ where $V_\text{k}$ is the Keplerian velocity at the disc, at a field strength of
\begin{equation}
\begin{aligned}
B_\text{WD}
&\leq \sqrt{8\pi \rho V_\text{k}^2} \\
&\leq 170\,\text{MG}
\left(\frac{\dot{M}}{10^{-5} M_{\odot}\,\text{yr}^{-1}}\right)^{1/2}
\left(\frac{M_{\text{WD}}}{0.6 M_{\odot}}\right)^{1/4} \\
&\quad \times
\left(\frac{\alpha_{\text{ss}}}{0.1}\right)^{-1/2}
\left(\frac{H/R}{0.05}\right)^{-3/2}
\left(\frac{R}{10^9\,\text{cm}}\right)^{-5/4}.
\end{aligned}
\label{Pres-balanced}
\end{equation}
On the other hand, 
the magnetic field strength advected over a viscous accretion time, when transported at a rate determined by free fall time,  or the Keplerian orbital frequency $\Omega_\text{k}$ at the disc's inner radius $\sim R_{\text{WD}}$, would be
\begin{equation}
    B_{\text{WD}}\leq  B_{d}\Omega_\text{k} (R_{\text{WD}}) t_{\text{v}}.
    \label{visc_field_exp}
\end{equation}
To compare equations (\ref{Pres-balanced}) and (\ref{visc_field_exp}), 
we need expressions for $B_\text{d}$ and $t_\text{v}$.
From equation (\ref{toroidal_field}), we have  $B_\text{d}\geq 8.94\sqrt{\pi \rho \alpha_\text{ss}} c_\text{s}$.

The viscous timescale 
is given by $t_{\text{v}} = \frac{R^2}{\nu_T}\simeq\frac{R_{\text{WD}}^2}{\alpha_\text{ss} c_\text{s} H}$,
where we have used the effective turbulent viscosity  $\nu_T=\alpha_{ss}c_\text{s} H$, 
and  $R\sim R_{WD}$ as the disc radius where the time scale is evaluated.
Combining these into equation (\ref{visc_field_exp}), the estimated total field accumulated viscously in $t_{\text{v}}$ would be
\begin{equation}
\begin{aligned}
B_{\rm WD}(t_v)
& \leq 15.87\sqrt{\rho V_k^2}\,
\frac{R_{\rm WD}}{\sqrt{\alpha_{\rm ss}}H} \\
&\leq 3.4\times10^{4}\,\mathrm{MG}
\left(\frac{\dot{M}}{10^{-5}M_\odot\,\mathrm{yr}^{-1}}\right)^{1/2}
\left(\frac{M_{\rm WD}}{0.6M_\odot}\right)^{1/4} \\
&\quad \times
\left(\frac{\alpha_{\rm ss}}{0.1}\right)^{-1}
\left(\frac{H/R}{0.05}\right)^{-5/2}
\left(\frac{R}{10^9\,\mathrm{cm}}\right)^{-5/4}.
\end{aligned}
\end{equation}
The inequality \(\frac{R_{\text{WD}}}{\sqrt{\alpha_\text{ss}} H} > 1\) always holds, which implies that the magnetic field accumulated always exceeds the pressure-limited threshold of equation (\ref{Pres-balanced}). Therefore, the  pressure balance limits the WD magnetic field to the value given by equation (\ref{Pres-balanced}).

\begin{figure*}[ht]
    \centering
    \begin{subfigure}[t]{0.48\textwidth}
        \centering
        \includegraphics[width=\linewidth]{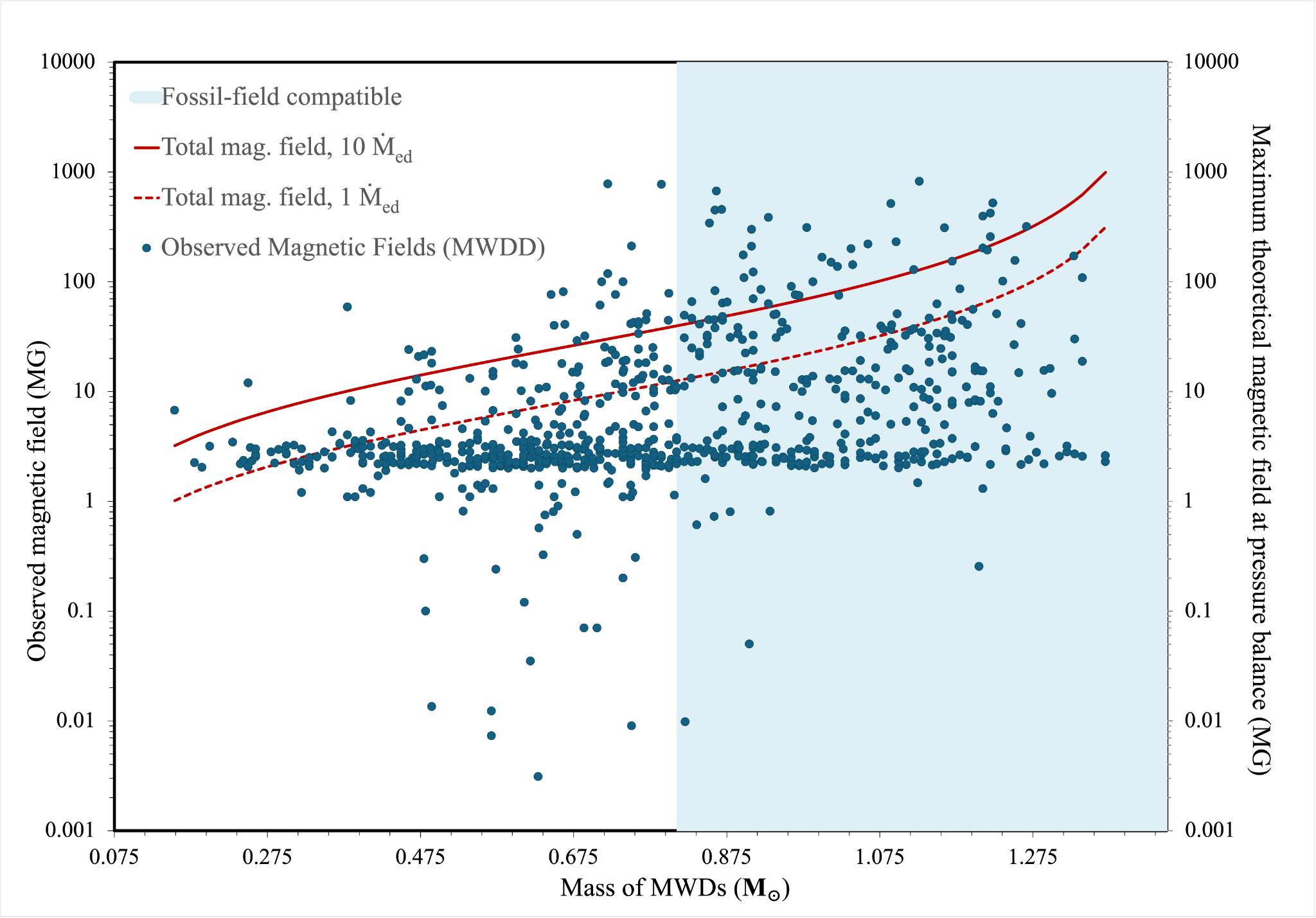}
        \caption{Full MWD sample.}
        \label{fig:MagField_Plim}
    \end{subfigure}
    \hfill
    \begin{subfigure}[t]{0.48\textwidth}
        \centering
        \includegraphics[width=\linewidth]{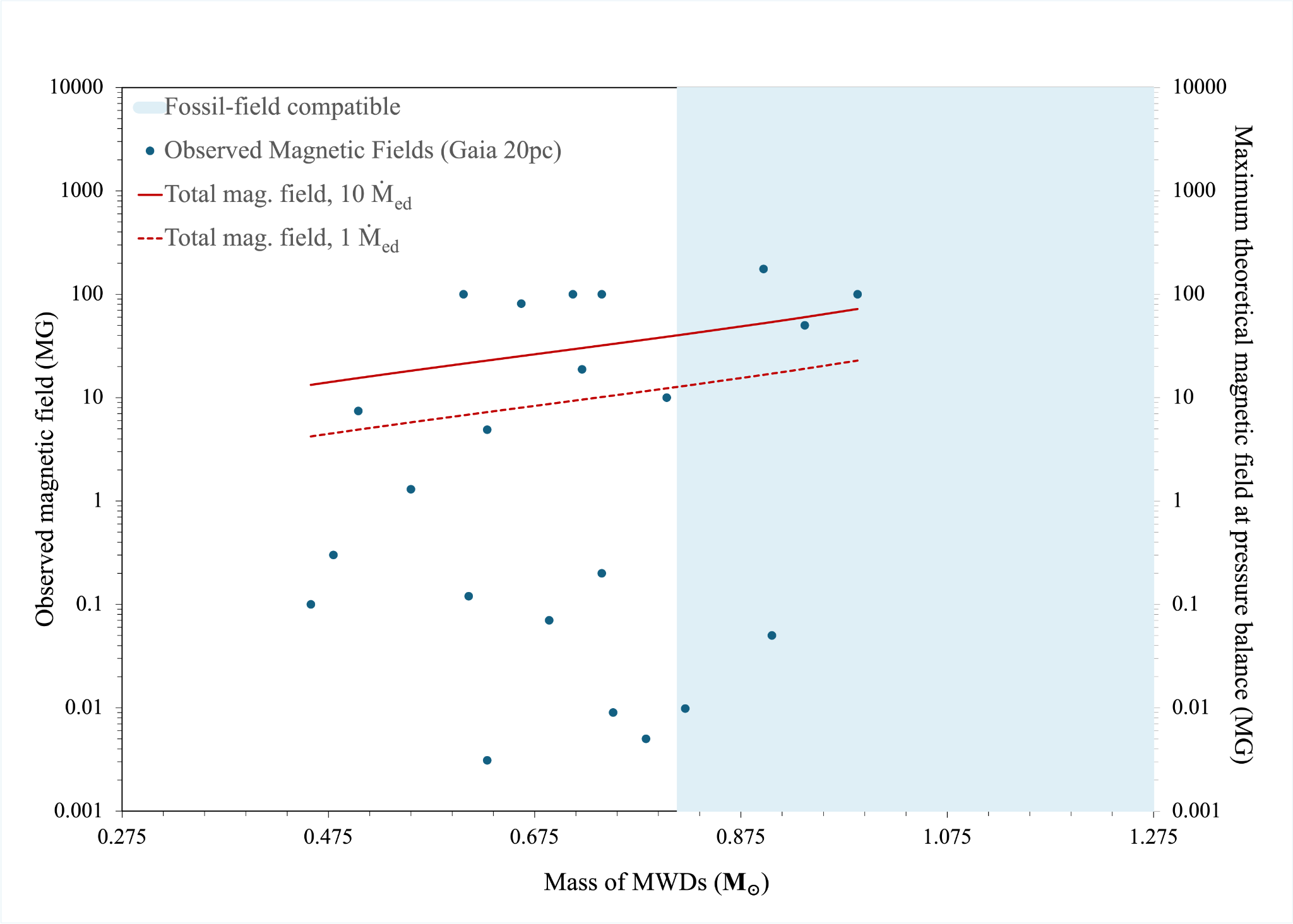}
        \caption{Volume-limited 20 pc sample.}
        \label{fig:MagField_Plim_VL}
    \end{subfigure}
    \caption{\textbf{Comparison of observed and theoretical white dwarf magnetic fields.}
    \textbf{(a)} Full MWD sample. \textbf{(b)} Volume-limited 20 pc sample. 
    The panels show observed magnetic field strengths (y-axis) as a function of mass (x-axis) for magnetic white dwarfs. 
    Red curves denote the theoretical upper limits of the total magnetic field strength, derived from pressure equilibrium in the accretion scenario for different accretion rates. The blue shaded region marks the mass range compatible with the fossil-field containment criterion.}
    \label{fig:MagFieldComparison}
\end{figure*}

As emphasised earlier, the observational methods often provide a lower limit to the total magnetic field strength estimated theoretically. In cases where the toroidal magnetic field is undetectable, the reported observed field strengths may more closely reflect  the net signed poloidal field.
\rev{In Figures \ref{fig:MagField_Plim} and \ref{fig:MagField_Plim_VL}, we compare the observationally reported MWD fields to the maximum total WD magnetic fields predicted by pressure equilibrium (red curves).} To establish the pressure-limited magnetic field value, we use equations (\ref{2a}),  (\ref{Pres-balanced}) and $V_\text{k} = \sqrt{\frac{G M}{R}}$ at  $R \sim R_{\text{WD}}$. We set $\dot M$ onto the WD to be a multiple of the Eddington rate of $\dot{M}_\text{ed} =   L_\text{ed}/ V_\text{k}^2 
= 10^{-5} M_\odot \text{yr}^{-1}$ 
for $R_\text{WD}=10^9$cm, $H/R = 0.447$ and $\alpha_\text{ss}=0.1$; any smaller $\alpha$ would produce even higher lines, making this choice a conservative one.  The dependence on $1/\alpha_{ss}^{1/2}$ for the lines corresponding to the total field arises from equation (\ref{2a}),
since we choose a fixed accretion rate. 
The dashed line in Figure \ref{fig:MagFieldComparison} corresponds to \(\dot{M} = \dot{M}_\text{ed}\) and the solid line corresponds to \(\dot{M} = 10 \dot{M}_\text{ed}\). \rev{Figure \ref{fig:MagField_Plim} shows a tendency for stronger magnetic fields ($> 100 \text{ MG}$) to be found in higher-mass MWDs including objects in the blue shaded fossil-field compatible region. This suggests that both fossil-field inheritance and accretion related amplification are plausible contributors to such high-mass, high-field MWDs. While lower-field MWDs are distributed over a wider range of masses, there is a noticeable clumping of objects around $B \sim 2-3 \text{ MG}$ which may reflect a combination of a higher occurrence of MWDs in that field range and observational selection effects. The lower edge of this clumping may be indicative of spectroscopic selection effects, since fields below $~1 \text{ MG}$ are harder to identify from Zeeman splitting \citep{2016A&A...591A..80L, 2026ApJ..1001..202A}. The decline above this range is not easily explained by any particular selection effect and may instead reflect a true decrease in occurrence rate of MWDs at higher field strengths.}


We do not expect accretion from a low-mass companion to strongly influence the WD temperature evolution. Because tidal disruption and accretion are expected to occur during the common-envelope phase, just before the very early stages of a young WD are exposed, we would expect that WDs which acquired fields this way would do so near the beginning of their cooling sequence. More detailed modelling of the timing of such events is required.

However, questions arise when evaluating whether  accretion of a very low-mass stellar or planetary companion can explain the majority of the MWD population, specifically, the event frequency and  the observed mass and age-distributions. Although a tidally shredded planetary companion can generate strong magnetic fields through accretion, it adds little to the mass of the final WD.
Magnetic fields in WDs appear to emerge at different stages in their evolution depending on mass. \rev{Polarimetric and spectroscopic studies have found that the strongest magnetic fields, typically $B>100 \text{ MG}$, are found in high-mass WDs almost immediately after their formation, whereas magnetism in normal-mass WDs appears only after 1-2 Gyr of cooling \citep{Bagnulo_2022,2025ApJ...990...25M}. We examine our compiled data  and find evidence for this trend, though the correlation is weak and shows substantial scatter: Figure \ref{fig:WDpercent_br} shows that above $T_{\mathrm{eff}} \sim 15000 \mathrm{K}$, most of the high-field $B \gtrsim 100 \text{ MG}$ WDs have $M \gtrsim  0.9 M_\odot$, whereas below $T_{\mathrm{eff}} \sim 15000 \mathrm{K}$ the mass distribution extends more broadly down to $M \gtrsim  0.5 M_\odot$.  The existence of the two lower mass objects above $15000 \mathrm{K}$ in this figure does show that such masses are not absolutely absent from the hotter population.}
As a result, to explain the WD mass and field strength correlation, MWDs produced through planet or low-mass stellar companion accretion would require the progenitor star to already have been massive.

\rev{An additional consideration for this channel is atmospheric pollution. If the tidally shredded companion is planetary and accretion is ongoing, significant metal features may be expected, as in DAZ/DZH systems. The predominance of DAH atmospheres among MWDs therefore disfavours recent planetary debris accretion as the dominant channel. Metals remain observable only during accretion and for a subsequent sinking time, so the visibility window is $t_\text{visible}\sim t_\text{acc}+t_\text{sink}$. Characteristic debris-accretion timescales are $\sim10^5\mathrm{yr}$ \citep{2009A&A...498..517K}, while metal sinking/settling times are $\sim10^4$--$10^6\mathrm{yr}$ \citep{2009ApJ...699.1473J}. These timescales are much shorter than the Gyr cooling ages of many WDs, so older planetary accretion events would not necessarily retain observable metal features \citep{2018haex.bookE..14Z}.}


\section*{Magnetic fields from WD-WD mergers}
\label{sec6}

\rev{As we discuss in this section, if the mass versus field strength correlation discussed in the previous section continues to survive more data,  WD-WD mergers can explain both the field strength versus WD mass correlation and the early appearance of strong magnetic fields. Moreover, WD-WD mergers may be required to explain the MWDs that lie well above the low-mass companion accretion-limited magnetic field envelope values discussed in the previous section.} 
However, whether such mergers occur frequently enough remains an open question:  population synthesis models \citep{bagnulo2020} provide estimates, though the actual rates depend on mass ratios, binary evolution and orbital separations. Additional channels like wind-accelerated binary mergers \citep{10.1093/mnras/stx2335} may help explain higher rates.

WD mergers produce a remnant consisting of  a central WD and a surrounding hot disc or torus formed from approximately half of the disrupted secondary star \citep{Garcia2013}. WD-WD mergers and tidal shredding of low-mass companions both involve accretion discs and some similarities in  magnetic field generation,  but self-gravity is more important for WD-WD merger discs.
 WD-WD mergers involve strong shear flows and high densities, conditions that are favourable for efficient magnetic field amplification.
Analytic estimates of dynamo amplification using the output of hydrodynamic simulations suggest that resulting magnetic field strengths could range from $10 - 10^{4}$ MG \citep{Garcia-Berro+2012,2016NewAR..71....9F, 2016MNRAS.458..325K}.

Direct three-dimensional magnetohydrodynamic simulations by \citet{2024A&A...691A.179P} demonstrate that mergers of helium white dwarfs robustly amplify seed magnetic fields by several orders of magnitude. The magnetic field evolution proceeds in two stages. During the dynamical merger, mutual disruption of the WDs results in strong shear and turbulence that drives a small-scale dynamo which rapidly amplifies the magnetic field. Following this initial phase, sustained differential rotation over many dynamical times drives a large-scale dynamo to further amplify magnetic fields at a much slower rate. The  magnetic field  saturates at strengths of order $\sim 10^{4}$ MG. 
The magnetic fields produced by these simulations are predominantly 
azimuthal and  should be interpreted as an upper bound on the mean poloidal magnetic field in the merger remnant.

Observationally inferred fields are  associated with the large-scale poloidal magnetic field component. 
Consequently, even a modest degree of reorganisation of the merger-generated magnetic field into a coherent large-scale poloidal component is sufficient to account for the highest-field MWDs.
This also turns out to be consistent with the estimated WD magnetic fields, derived from pressure balance in an accretion scenario from  equation(\ref{Pres-balanced}), using plausible accretion rates $\dot{M}\leq 10 {\dot{M}}_\text{ed}$. While the accretion disc model for mergers requires self-gravity, this ram pressure may still establish a reasonable upper limit for the magnetic field. 

\rev{As discussed earlier in the crystallisation section, to reduce the observational  bias towards younger WDs present in the full MWD sample  we repeated our analyses with a 20 pc volume-limited sample from Gaia DR2. As shown in Figure \ref{fig:MagField_Plim_VL}, the observed magnetic field strengths of WDs in this more complete sub-sample are broadly consistent with the WD-WD merger and accretion limits from pressure balance. This supports the plausibility that both mergers and companion accretion can play significant respective roles encompassing a  broad range of MWDs. }

Finally, simulations indicate that merger remnants are heated to temperatures of order $10^7$ K, which is much higher than  typically assumed in standard cooling models. Such heating effectively resets the cooling clock, explaining why young cooling ages are inferred for many high field MWDs. 
Observationally, however, only massive MWDs have been reported to be young, whereas magnetic fields in normal-mass WDs, which constitute the dominant population, tend to emerge only after delays of $1-2$ Gyr \citep{Bagnulo_2022, 2025ApJ...990...25M}. This is not yet fully understood.  Possible explanations include (i) a lower fraction of merger products among the normal-mass WDs; (ii) delayed emergence of merger-generated magnetic fields due to initial burial beneath the surface; or (iii) reduced efficiency in sustaining large-scale poloidal fields in lower-mass merger remnants such that e.g. crystallisation dynamos play a larger role in that population.
\section{Conclusions}
\label{sec:discussion}

Our analysis constrains the viable proposed mechanisms for MWDs when examined against existing observational and theoretical constraints. 
\rev{Dynamo mechanisms that depend on companions cannot be 
ruled out as an explanation for the magnetic fields in most MWDs, as summarised in Figure \ref{fig:MagField_Plim}. On the basis of companion types, WD-WD mergers more naturally explain the young, high-mass, strongly magnetised MWDs since both WDs add significantly to the final WD mass and produce strong shear for magnetic-field amplification. Tidally shredded planetary or low-mass companions, on the other hand, add very little to the final mass and are more relevant to a subset of average-mass MWDs.}
The same trends are found for the full sample and a 
volume-limited sample from Gaia as shown in Figure \ref{fig:MagField_Plim_VL}.  
More detailed theoretical investigations of accretion and merger models are warranted to better understand the field generation and  advection of the field onto the WD core.  The position of new WDs on future versions of Figure \ref{fig:MagField_Plim} will help to further constrain the field origin from these mechanisms. 

\rev{We also showed that the fossil field origin is difficult to rule out at present. The problem here is the overproduction of MWDs unless the field remains trapped within the WD or decays before convection fully shuts off.}


Our comparison of MWD cooling ages with the theoretically expected crystallisation onset ages reveals that crystallisation occurs too late to be the dominant mechanism for magnetic field generation in the majority of  MWDs.
The analysis does, however, reveal the  potential role of  crystallisation dynamos for older, cooler WDs, particularly
 MWDs in the 5000-7500 K range, and more data are needed to pin  down this subset of systems.
 
The incompleteness of current MWD data, and the need for more, is evidenced by the fact that the fraction of MWDs as a function of cooling age decreases with WD age for large sections of Figure \ref{fig:WDpercent}, which would not occur if the sample were complete. 

\begin{acknowledgement}
{\sloppy
We thank I. Caiazzo, J. Nordhaus, L. Chamandy, and S. Ginzburg for helpful discussions. We acknowledge partial support from the National Science Foundation (NSF) under award no. PHY-2020249 and from the Aspen Center for Physics, which is supported by the NSF under award no. PHY-2210452. We acknowledge support from the U.S. DOE NNSA under award nos. DE-NA0004144 and DE-NA0004147 and through subcontracts nos. 630138 and C4574 with Los Alamos National Laboratory to the Flash Center for Computational Science. Support from the U.S. DOE Office of Science, Fusion Energy Sciences, under award no. DE-SC0021990 is also acknowledged.
\par}
\end{acknowledgement}


\bibliography{mwd}

@ARTICLE{Braithwaite+2004,
       author = {{Braithwaite}, Jonathan and {Spruit}, Hendrik C.},
        title = "{A fossil origin for the magnetic field in A stars and white dwarfs}",
      journal = {\nat},
         year = 2004,
        month = oct,
       volume = {431},
       number = {7010},
        pages = {819-821},
          doi = {10.1038/nature02934},
archivePrefix = {arXiv},
       eprint = {astro-ph/0502043},
 primaryClass = {astro-ph},
       adsurl = {https://ui.adsabs.harvard.edu/abs/2004Natur.431..819B}
}

@ARTICLE{Ginzburg1964,
       author = {{Ginzburg}, V.~L.},
        title = "{The Magnetic Fields of Collapsing Masses and the Nature of Superstars}",
      journal = {Soviet Physics Doklady},
         year = 1964,
        month = nov,
       volume = {9},
        pages = {329},
       adsurl = {https://ui.adsabs.harvard.edu/abs/1964SPhD....9..329G}
}

@ARTICLE{2016MNRAS.458..325K,
       author = {{Kawka}, Adela and {Vennes}, St{\'e}phane},
        title = "{Extreme abundance ratios in the polluted atmosphere of the cool white dwarf NLTT 19868}",
      journal = {\mnras},
         year = 2016,
        month = may,
       volume = {458},
       number = {1},
        pages = {325-331},
          doi = {10.1093/mnras/stw383},
archivePrefix = {arXiv},
       eprint = {1602.05000},
 primaryClass = {astro-ph.SR},
       adsurl = {https://ui.adsabs.harvard.edu/abs/2016MNRAS.458..325K}
}

@ARTICLE{Woltjer1964,
       author = {{Woltjer}, L.},
        title = "{X-Rays and Type I Supernova Remnants.}",
      journal = {\apj},
         year = 1964,
        month = oct,
       volume = {140},
        pages = {1309-1313},
          doi = {10.1086/148028},
       adsurl = {https://ui.adsabs.harvard.edu/abs/1964ApJ...140.1309W}
}

@ARTICLE{2016NewAR..71....9F,
       author = {{Farihi}, J.},
        title = "{Circumstellar debris and pollution at white dwarf stars}",
      journal = {\nar},
         year = 2016,
        month = apr,
       volume = {71},
        pages = {9-34},
          doi = {10.1016/j.newar.2016.03.001},
archivePrefix = {arXiv},
       eprint = {1604.03092},
 primaryClass = {astro-ph.EP},
       adsurl = {https://ui.adsabs.harvard.edu/abs/2016NewAR..71....9F}
}

@ARTICLE{2009ApJ...699.1473J,
       author = {{Jura}, M. and {Muno}, M.~P. and {Farihi}, J. and {Zuckerman}, B.},
        title = "{X-Ray and Infrared Observations of Two Externally Polluted White Dwarfs}",
      journal = {\apj},
         year = 2009,
        month = jul,
       volume = {699},
       number = {2},
        pages = {1473-1479},
          doi = {10.1088/0004-637X/699/2/1473},
archivePrefix = {arXiv},
       eprint = {0905.0117},
 primaryClass = {astro-ph.SR},
       adsurl = {https://ui.adsabs.harvard.edu/abs/2009ApJ...699.1473J}
}

@ARTICLE{2009A&A...498..517K,
       author = {{Koester}, D.},
        title = "{Accretion and diffusion in white dwarfs. New diffusion timescales and applications to GD 362 and G 29-38}",
      journal = {\aap},
         year = 2009,
        month = may,
       volume = {498},
       number = {2},
        pages = {517-525},
          doi = {10.1051/0004-6361/200811468},
archivePrefix = {arXiv},
       eprint = {0903.1499},
 primaryClass = {astro-ph.SR},
       adsurl = {https://ui.adsabs.harvard.edu/abs/2009A&A...498..517K}
}

@INCOLLECTION{2018haex.bookE..14Z,
       author = {{Zuckerman}, B. and {Young}, E.~D.},
        title = "{Characterizing the Chemistry of Planetary Materials Around White Dwarf Stars}",
    booktitle = {Handbook of Exoplanets},
         year = 2018,
       editor = {{Deeg}, Hans J. and {Belmonte}, Juan Antonio},
          eid = {14},
        pages = {14},
          doi = {10.1007/978-3-319-55333-7_14},
       adsurl = {https://ui.adsabs.harvard.edu/abs/2018haex.bookE..14Z}
}

@ARTICLE{2024ApJ...961..197M,
       author = {{Montgomery}, M.~H. and {Dunlap}, Bart H.},
        title = "{Fluid Mixing during Phase Separation in Crystallizing White Dwarfs}",
      journal = {\apj},
         year = 2024,
        month = feb,
       volume = {961},
       number = {2},
          eid = {197},
        pages = {197},
          doi = {10.3847/1538-4357/ad16dc},
archivePrefix = {arXiv},
       eprint = {2312.11647},
 primaryClass = {astro-ph.SR},
       adsurl = {https://ui.adsabs.harvard.edu/abs/2024ApJ...961..197M}
}

@ARTICLE{2026A&A...708L..14E,
       author = {{Einramhof}, L. and {Bugnet}, L. and {Calcaferro}, L.~M. and {Barrault}, L. and {Das}, S.~B.},
        title = "{Magneto-archeology of white dwarfs: Revisiting the fossil field scenario with observational constraints during the red giant branch}",
      journal = {\aap},
         year = 2026,
        month = apr,
       volume = {708},
          eid = {L14},
        pages = {L14},
          doi = {10.1051/0004-6361/202659069},
archivePrefix = {arXiv},
       eprint = {2601.15203},
 primaryClass = {astro-ph.SR},
       adsurl = {https://ui.adsabs.harvard.edu/abs/2026A&A...708L..14E}
}

@ARTICLE{2016A&A...591A..80L,
       author = {{Landstreet}, J.~D. and {Bagnulo}, S. and {Martin}, A. and {Valyavin}, G.},
        title = "{Discovery of an extremely weak magnetic field in the white dwarf LTT 16093 = WD 2047+372}",
      journal = {\aap},
         year = 2016,
        month = jun,
       volume = {591},
          eid = {A80},
        pages = {A80},
          doi = {10.1051/0004-6361/201628488},
archivePrefix = {arXiv},
       eprint = {1605.04458},
 primaryClass = {astro-ph.SR},
       adsurl = {https://ui.adsabs.harvard.edu/abs/2016A&A...591A..80L}
}

@ARTICLE{2026ApJ..1001..202A,
       author = {{Amorim}, L.~L. and {Kepler}, S.~O. and {Romero}, Alejandra D.},
        title = "{Magnetic White Dwarfs from DESI}",
      journal = {\apj},
         year = 2026,
        month = apr,
       volume = {1001},
       number = {2},
          eid = {202},
        pages = {202},
          doi = {10.3847/1538-4357/ae5524},
archivePrefix = {arXiv},
       eprint = {2603.20487},
 primaryClass = {astro-ph.SR},
       adsurl = {https://ui.adsabs.harvard.edu/abs/2026ApJ..1001..202A}
}

@ARTICLE{2023ApJ...944...56A,
       author = {{Amorim}, L.~L. and {Kepler}, S.~O. and {K{\"u}lebi}, Baybars and {Jordan}, S. and {Romero}, A.~D.},
        title = "{Catalog of Magnetic White Dwarfs with Hydrogen Dominated Atmospheres}",
      journal = {\apj},
         year = 2023,
        month = feb,
       volume = {944},
       number = {1},
          eid = {56},
        pages = {56},
          doi = {10.3847/1538-4357/acaf6e},
archivePrefix = {arXiv},
       eprint = {2301.08862},
 primaryClass = {astro-ph.SR},
       adsurl = {https://ui.adsabs.harvard.edu/abs/2023ApJ...944...56A}
}

@article{Castro-Tapia_2026,
doi = {10.3847/1538-4357/ae2fb6},
url = {https://doi.org/10.3847/1538-4357/ae2fb6},
year = {2026},
month = {feb},
publisher = {The American Astronomical Society},
volume = {998},
number = {1},
pages = {28},
author = {Castro-Tapia, Matias and Camisassa, Maria and Zhang, Shu},
title = {Evolution of a Long-lived Deep-seated Main-sequence Magnetic Field during White Dwarf Cooling},
journal = {The Astrophysical Journal}
}

@ARTICLE{2025ApJ...990...25M,
       author = {{Moss}, Adam and {Kilic}, Mukremin and {Bergeron}, Pierre and {Brown}, Warren R. and {Jewett}, Gracyn and {Ag{\"u}eros}, Marcel A. and {Camisassa}, Maria and {Burrow}, Anthony},
        title = "{Magnetic White Dwarfs in the SDSS 100 pc Sample: Further Evidence of Two Formation Channels}",
      journal = {\apj},
         year = 2025,
        month = sep,
       volume = {990},
       number = {1},
          eid = {25},
        pages = {25},
          doi = {10.3847/1538-4357/aded8f},
archivePrefix = {arXiv},
       eprint = {2507.06102},
 primaryClass = {astro-ph.SR},
       adsurl = {https://ui.adsabs.harvard.edu/abs/2025ApJ...990...25M}
}

@BOOK{2013sse..book.....K,
       author = {{Kippenhahn}, Rudolf and {Weigert}, Alfred and {Weiss}, Achim},
        title = "{Stellar Structure and Evolution}",
         year = 2013,
          doi = {10.1007/978-3-642-30304-3},
       adsurl = {https://ui.adsabs.harvard.edu/abs/2013sse..book.....K}
}

@ARTICLE{1997A&A...327.1039C,
       author = {{Chabrier}, Gilles and {Baraffe}, Isabelle},
        title = "{Structure and evolution of low-mass stars}",
      journal = {\aap},
         year = 1997,
        month = nov,
       volume = {327},
        pages = {1039-1053},
          doi = {10.48550/arXiv.astro-ph/9704118},
archivePrefix = {arXiv},
       eprint = {astro-ph/9704118},
 primaryClass = {astro-ph},
       adsurl = {https://ui.adsabs.harvard.edu/abs/1997A&A...327.1039C}
}

@ARTICLE{2024A&A...691A.179P,
       author = {{Pakmor}, R{\"u}diger and {Pelisoli}, Ingrid and {Justham}, Stephen and {Rajamuthukumar}, Abinaya S. and {R{\"o}pke}, Friedrich K. and {Schneider}, Fabian R.~N. and {de Mink}, Selma E. and {Ohlmann}, Sebastian T. and {Podsiadlowski}, Philipp and {Mor{\'a}n-Fraile}, Javier and {Vetter}, Marco and {Andrassy}, Robert},
        title = "{Large-scale ordered magnetic fields generated in mergers of helium white dwarfs}",
      journal = {\aap},
         year = 2024,
        month = nov,
       volume = {691},
          eid = {A179},
        pages = {A179},
          doi = {10.1051/0004-6361/202451352},
archivePrefix = {arXiv},
       eprint = {2407.02566},
 primaryClass = {astro-ph.SR},
       adsurl = {https://ui.adsabs.harvard.edu/abs/2024A&A...691A.179P}
}

@ARTICLE{Zhu20132013ApJ...767..164Z,
       author = {{Zhu}, Chenchong and {Chang}, Philip and {van Kerkwijk}, Marten H. and {Wadsley}, James},
        title = "{A Parameter-space Study of Carbon-Oxygen White Dwarf Mergers}",
      journal = {\apj},
         year = 2013,
        month = apr,
       volume = {767},
       number = {2},
          eid = {164},
        pages = {164},
          doi = {10.1088/0004-637X/767/2/164},
archivePrefix = {arXiv},
       eprint = {1210.3616},
 primaryClass = {astro-ph.SR},
       adsurl = {https://ui.adsabs.harvard.edu/abs/2013ApJ...767..164Z}
}

@article{Castro_Tapia_2024,
   title={Magnetic Field Evolution for Crystallization-driven Dynamos in C/O White Dwarfs},
   volume={975},
   ISSN={1538-4357},
   url={http://dx.doi.org/10.3847/1538-4357/ad7a6a},
   DOI={10.3847/1538-4357/ad7a6a},
   number={1},
   journal={\apj},
   publisher={American Astronomical Society},
   author={Castro-Tapia, Matias and Zhang, Shu and Cumming, Andrew},
   year={2024},
   month=oct, pages={63} }

@article{ SeismoGangLi,
	author = {{Li, Gang} and {Deheuvels, Sébastien} and {Li, Tanda} and {Ballot, Jérôme} and {Lignières, François}},
	title = {Internal magnetic fields in 13 red giants detected by asteroseismology},
	DOI= "10.1051/0004-6361/202347260",
	url= "https://doi.org/10.1051/0004-6361/202347260",
	journal = {Astronomy and Astrophysics},
	year = 2023,
	volume = 680,
	pages = "A26",
}

@article{Fuller_2015,
   title={Asteroseismology can reveal strong internal magnetic fields in red giant stars},
   volume={350},
   ISSN={1095-9203},
   url={http://dx.doi.org/10.1126/science.aac6933},
   DOI={10.1126/science.aac6933},
   number={6259},
   journal={Science},
   publisher={American Association for the Advancement of Science (AAAS)},
   author={Fuller, Jim and Cantiello, Matteo and Stello, Dennis and Garcia, Rafael A. and Bildsten, Lars},
   year={2015},
   month=oct, pages={423–426} }

@article{ChrisScaling,
    author  = {Christensen, U. R. and Aubert, J.},
    title = {Scaling properties of convection-driven dynamos in rotating spherical shells and application to planetary magnetic fields},
    journal = {Geophysical Journal International},
    volume = {166},
    number = {1},
    pages = {97-114},
    year = {2006},
    month = {07},
    issn = {0956-540X},
    doi = {10.1111/j.1365-246X.2006.03009.x},
    url = {https://doi.org/10.1111/j.1365-246X.2006.03009.x},
    eprint = {https://academic.oup.com/gji/article-pdf/166/1/97/5891407/166-1-97.pdf},
}

@article{Camisassa2024,
    author = {Camisassa, M. and Fuentes, J. R. and Schreiber, M. R. and Rebassa-Mansergas, A. and Torres, S. and Raddi, R. and Dominguez, I.},
    title = {Main sequence dynamo magnetic fields emerging in the white dwarf phase},
    DOI = "10.1051/0004-6361/202452539",
    url = "https://doi.org/10.1051/0004-6361/202452539",
    journal = {Astronomy and Astrophysics},
    year = 2024,
    volume = 691,
    pages = "L21",
}

@article{10.1093/mnras/sty2057,
    author = {Hollands, M A and Tremblay, P-E and Gänsicke, B T and Gentile-Fusillo, N P and Toonen, S},
    title = {The Gaia 20pc white dwarf sample},
    journal = {Monthly Notices of the Royal Astronomical Society},
    volume = {480},
    number = {3},
    pages = {3942-3961},
    year = {2018},
    month = {08},
    issn = {0035-8711},
    doi = {10.1093/mnras/sty2057},
    url = {https://doi.org/10.1093/mnras/sty2057},
    eprint = {https://academic.oup.com/mnras/article-pdf/480/3/3942/25524012/sty2057.pdf},
}

@article{10.1093/mnras/stx2335,
    author = {Chen, Zhuo and Blackman, Eric G. and Nordhaus, Jason and Frank, Adam and Carroll-Nellenback, Jonathan},
    title = {Wind-accelerated orbital evolution in binary systems with giant stars},
    journal = {Monthly Notices of the Royal Astronomical Society},
    volume = {473},
    number = {1},
    pages = {747-756},
    year = {2017},
    month = {09},
    issn = {0035-8711},
    doi = {10.1093/mnras/stx2335},
    url = {https://doi.org/10.1093/mnras/stx2335},
    eprint = {https://academic.oup.com/mnras/article-pdf/473/1/747/21370681/stx2335.pdf},
}

@article{ Toonen,
	author = {{Toonen, S.} and {Hollands, M.} and {Gänsicke, B. T.} and {Boekholt, T.}},
	title = {The binarity of the local white dwarf population },
	DOI= "10.1051/0004-6361/201629978",
	url= "https://doi.org/10.1051/0004-6361/201629978",
	journal = {Astronomy and Astrophysics},
	year = 2017,
	volume = 602,
	pages = "A16",
}

@ARTICLE{2008A&A...481..465L,
       author = {{Landstreet}, J.~D. and {Silaj}, J. and {Andretta}, V. and {Bagnulo}, S. and {Berdyugina}, S.~V. and {Donati}, J. -F. and {Fossati}, L. and {Petit}, P. and {Silvester}, J. and {Wade}, G.~A.},
        title = "{Searching for links between magnetic fields and stellar evolution. III. Measurement of magnetic fields in open cluster Ap stars with ESPaDOnS}",
      journal = {Astronomy and Astrophysics},
         year = 2008,
        month = apr,
       volume = {481},
       number = {2},
        pages = {465-480},
          doi = {10.1051/0004-6361:20078884},
archivePrefix = {arXiv},
       eprint = {0803.0877},
 primaryClass = {astro-ph},
       adsurl = {https://ui.adsabs.harvard.edu/abs/2008A&A...481..465L}
}

@article{Kirzhnits1960,
  title={Internal structure of super-dense stars},
  author={Kirzhnits, DA},
  journal={Sov. Phys. JETP},
  volume={11},
  pages={365--368},
  year={1960}
}

@article{Stevenson1980,
  title={A eutectic in carbon-oxygen white dwarfs?},
  author={Stevenson, DJ},
  journal={Le Journal de Physique Colloques},
  volume={41},
  number={C2},
  pages={C2--61},
  year={1980},
  publisher={EDP Sciences}
}

@article{VanHorn1968,
  title={Crystallization of white dwarfs},
  author={Van Horn, HM},
  journal={Astrophysical Journal, vol. 151, p. 227},
  volume={151},
  pages={227},
  year={1968}
}

@book{Frank+2002, place={Cambridge}, edition={3}, title={Accretion Power in Astrophysics}, publisher={Cambridge University Press}, author={Frank, Juhan and King, Andrew and Raine, Derek}, year={2002}}

@ARTICLE{Guidarelli+19,
       author = {{Guidarelli}, G. and {Nordhaus}, J. and {Chamandy}, L. and {Chen}, Z. and
         {Blackman}, E.~G. and {Frank}, A. and {Carroll-Nellenback}, J. and
         {Liu}, B.},
        title = "{Hydrodynamic simulations of disrupted planetary accretion discs inside the core of an AGB star}",
      journal = {\mnras},
         year = 2019,
        month = nov,
       volume = {490},
       number = {1},
        pages = {1179-1185},
          doi = {10.1093/mnras/stz2641},
archivePrefix = {arXiv},
       eprint = {1908.00157},
 primaryClass = {astro-ph.SR},
       adsurl = {https://ui.adsabs.harvard.edu/abs/2019MNRAS.490.1179G}
}

@ARTICLE{Blackman+2008,
       author = {{Blackman}, E.~G. and {Penna}, R.~F. and {Varni{\`e}re}, P.},
        title = "{Empirical relation between angular momentum transport and thermal-to-magnetic pressure ratio in shearing box simulations}",
      journal = {\na},
         year = 2008,
        month = may,
       volume = {13},
       number = {4},
        pages = {244-251},
          doi = {10.1016/j.newast.2007.10.004},
archivePrefix = {arXiv},
       eprint = {astro-ph/0607119},
 primaryClass = {astro-ph},
       adsurl = {https://ui.adsabs.harvard.edu/abs/2008NewA...13..244B}
}

@ARTICLE{Kiuchi+2024,
       author = {{Kiuchi}, Kenta and {Reboul-Salze}, Alexis and {Shibata}, Masaru and {Sekiguchi}, Yuichiro},
        title = "{A large-scale magnetic field produced by a solar-like dynamo in binary neutron star mergers}",
      journal = {Nature Astronomy},
         year = 2024,
        month = mar,
       volume = {8},
        pages = {298-307},
          doi = {10.1038/s41550-024-02194-y},
archivePrefix = {arXiv},
       eprint = {2306.15721},
 primaryClass = {astro-ph.HE},
       adsurl = {https://ui.adsabs.harvard.edu/abs/2024NatAs...8..298K}
}

@ARTICLE{Ondratschek+2022,
       author = {{Ondratschek}, Patrick A. and {R{\"o}pke}, Friedrich K. and {Schneider}, Fabian R.~N. and {Fendt}, Christian and {Sand}, Christian and {Ohlmann}, Sebastian T. and {Pakmor}, R{\"u}diger and {Springel}, Volker},
        title = "{Bipolar planetary nebulae from common-envelope evolution of binary stars}",
      journal = {Astronomy and Astrophysics},
         year = 2022,
        month = apr,
       volume = {660},
          eid = {L8},
        pages = {L8},
          doi = {10.1051/0004-6361/202142478},
archivePrefix = {arXiv},
       eprint = {2110.13177},
 primaryClass = {astro-ph.SR},
       adsurl = {https://ui.adsabs.harvard.edu/abs/2022A&A...660L...8O}
}

@ARTICLE{Nordhaus+2006,
       author = {{Nordhaus}, J. and {Blackman}, E.~G.},
        title = "{Low-mass binary-induced outflows from asymptotic giant branch stars}",
      journal = {\mnras},
         year = 2006,
        month = aug,
       volume = {370},
       number = {4},
        pages = {2004-2012},
          doi = {10.1111/j.1365-2966.2006.10625.x},
archivePrefix = {arXiv},
       eprint = {astro-ph/0604445},
 primaryClass = {astro-ph},
       adsurl = {https://ui.adsabs.harvard.edu/abs/2006MNRAS.370.2004N}
}

@ARTICLE{Kissin+2015,
       author = {{Kissin}, Yevgeni and {Thompson}, Christopher},
        title = "{Spin and Magnetism of White Dwarfs}",
      journal = {\apj},
         year = 2015,
        month = aug,
       volume = {809},
       number = {2},
          eid = {108},
        pages = {108},
          doi = {10.1088/0004-637X/809/2/108},
archivePrefix = {arXiv},
       eprint = {1501.07197},
 primaryClass = {astro-ph.SR},
       adsurl = {https://ui.adsabs.harvard.edu/abs/2015ApJ...809..108K}
}

@ARTICLE{Regos+1995,
       author = {{Reg{\H{o}}s}, Enik{\H{o}} and {Tout}, Christopher A.},
        title = "{The effect of magnetic fields in common-envelope evolution on the formation of cataclysmic variables}",
      journal = {\mnras},
         year = 1995,
        month = mar,
       volume = {273},
       number = {1},
        pages = {146-156},
          doi = {10.1093/mnras/273.1.146},
       adsurl = {https://ui.adsabs.harvard.edu/abs/1995MNRAS.273..146R}
}

@ARTICLE{Kepler+2013,
       author = {{Kepler}, S.~O. and {Pelisoli}, I. and {Jordan}, S. and {Kleinman}, S.~J. and {Koester}, D. and {K{\"u}lebi}, B. and {Pe{\c{c}}anha}, V. and {Castanheira}, B.~G. and {Nitta}, A. and {Costa}, J.~E.~S. and {Winget}, D.~E. and {Kanaan}, A. and {Fraga}, L.},
        title = "{Magnetic white dwarf stars in the Sloan Digital Sky Survey}",
      journal = {\mnras},
         year = 2013,
        month = mar,
       volume = {429},
       number = {4},
        pages = {2934-2944},
          doi = {10.1093/mnras/sts522},
archivePrefix = {arXiv},
       eprint = {1211.5709},
 primaryClass = {astro-ph.SR},
       adsurl = {https://ui.adsabs.harvard.edu/abs/2013MNRAS.429.2934K}
}

@ARTICLE{Tout+2008,
       author = {{Tout}, C.~A. and {Wickramasinghe}, D.~T. and {Liebert}, J. and {Ferrario}, L. and {Pringle}, J.~E.},
        title = "{Binary star origin of high field magnetic white dwarfs}",
      journal = {\mnras},
         year = 2008,
        month = jun,
       volume = {387},
       number = {2},
        pages = {897-901},
          doi = {10.1111/j.1365-2966.2008.13291.x},
archivePrefix = {arXiv},
       eprint = {0805.0115},
 primaryClass = {astro-ph},
       adsurl = {https://ui.adsabs.harvard.edu/abs/2008MNRAS.387..897T}
}

@ARTICLE{Nordhaus+2007,
       author = {{Nordhaus}, J. and {Blackman}, E.~G. and {Frank}, A.},
        title = "{Isolated versus common envelope dynamos in planetary nebula progenitors}",
      journal = {\mnras},
         year = 2007,
        month = apr,
       volume = {376},
       number = {2},
        pages = {599-608},
          doi = {10.1111/j.1365-2966.2007.11417.x},
archivePrefix = {arXiv},
       eprint = {astro-ph/0609726},
 primaryClass = {astro-ph},
       adsurl = {https://ui.adsabs.harvard.edu/abs/2007MNRAS.376..599N}
}

@article{10.1093/mnras/stt1910,
    author = {Wickramasinghe, Dayal T. and Tout, Christopher A. and Ferrario, Lilia},
    title = "{The most magnetic stars}",
    journal = {Monthly Notices of the Royal Astronomical Society},
    volume = {437},
    number = {1},
    pages = {675-681},
    year = {2013},
    month = {11},
    issn = {0035-8711},
    doi = {10.1093/mnras/stt1910},
    url = {https://doi.org/10.1093/mnras/stt1910},
    eprint = {https://academic.oup.com/mnras/article-pdf/437/1/675/18457709/stt1910.pdf},
}

@article{Briggs_2014,
   title={Merging binary stars and the magnetic white dwarfs},
   volume={447},
   ISSN={0035-8711},
   url={http://dx.doi.org/10.1093/mnras/stu2539},
   DOI={10.1093/mnras/stu2539},
   number={2},
   journal={Monthly Notices of the Royal Astronomical Society},
   publisher={Oxford University Press (OUP)},
   author={Briggs, Gordon P. and Ferrario, Lilia and Tout, Christopher A. and Wickramasinghe, Dayal T. and Hurley, Jarrod R.},
   year={2014},
   month=dec, pages={1713–1723} }

@ARTICLE{2005MNRAS.356..615F,
       author = {{Ferrario}, Lilia and {Wickramasinghe}, D.~T.},
        title = "{Magnetic fields and rotation in white dwarfs and neutron stars}",
      journal = {\mnras},
         year = 2005,
        month = jan,
       volume = {356},
       number = {2},
        pages = {615-620},
          doi = {10.1111/j.1365-2966.2004.08474.x},
       adsurl = {https://ui.adsabs.harvard.edu/abs/2005MNRAS.356..615F}
}

@article{Isern_2017,
   title={A Common Origin of Magnetism from Planets to White Dwarfs},
   volume={836},
   ISSN={2041-8213},
   url={http://dx.doi.org/10.3847/2041-8213/aa5eae},
   DOI={10.3847/2041-8213/aa5eae},
   number={2},
   journal={\apjl},
   publisher={American Astronomical Society},
   author={Isern, Jordi and García-Berro, Enrique and Külebi, Baybars and Lorén-Aguilar, Pablo},
   year={2017},
   month=feb, pages={L28} }

@article{Hardy_2023,
   title={Spectrophotometric analysis of magnetic white dwarf – I. Hydrogen-rich compositions},
   volume={520},
   ISSN={1365-2966},
   url={http://dx.doi.org/10.1093/mnras/stad196},
   DOI={10.1093/mnras/stad196},
   number={4},
   journal={Monthly Notices of the Royal Astronomical Society},
   publisher={Oxford University Press (OUP)},
   author={Hardy, François and Dufour, Patrick and Jordan, Stefan},
   year={2023},
   month=jan, pages={6111–6134} }

@article{Blackman_2001,
   title={Magnetohydrodynamic Stellar and Disk Winds: Application to Planetary Nebulae},
   volume={546},
   ISSN={1538-4357},
   url={http://dx.doi.org/10.1086/318253},
   DOI={10.1086/318253},
   number={1},
   journal={\apj},
   publisher={American Astronomical Society},
   author={Blackman, Eric G. and Frank, Adam and Welch, Carl},
   year={2001},
   month=jan, pages={288–298} }

@ARTICLE{Fuentes+2024,
       author = {{Fuentes}, J.~R. and {Castro-Tapia}, Matias and {Cumming}, Andrew},
        title = "{A Short Intense Dynamo at the Onset of Crystallization in White Dwarfs}",
      journal = {\apjl},
         year = 2024,
        month = mar,
       volume = {964},
       number = {1},
          eid = {L15},
        pages = {L15},
          doi = {10.3847/2041-8213/ad3100},
archivePrefix = {arXiv},
       eprint = {2402.03639},
 primaryClass = {astro-ph.SR},
       adsurl = {https://ui.adsabs.harvard.edu/abs/2024ApJ...964L..15F}
}

@article{Tremblay_2019,
   title={Core  crystallization and pile-up in the cooling sequence of evolving white dwarfs},
   volume={565},
   ISSN={1476-4687},
   url={http://dx.doi.org/10.1038/s41586-018-0791-x},
   DOI={10.1038/s41586-018-0791-x},
   number={7738},
   journal={Nature},
   publisher={Springer Science and Business Media LLC},
   author={Tremblay, Pier-Emmanuel and Fontaine, Gilles and Fusillo, Nicola Pietro Gentile and Dunlap, Bart H. and Gänsicke, Boris T. and Hollands, Mark A. and Hermes, J. J. and Marsh, Thomas R. and Cukanovaite, Elena and Cunningham, Tim},
   year={2019},
   month=jan, pages={202–205} }

@ARTICLE{2020ApJ...901...93B,
       author = {{B{\'e}dard}, A. and {Bergeron}, P. and {Brassard}, P. and {Fontaine}, G.},
        title = "{On the Spectral Evolution of Hot White Dwarf Stars. I. A Detailed Model Atmosphere Analysis of Hot White Dwarfs from SDSS DR12}",
      journal = {\apj},
         year = 2020,
        month = oct,
       volume = {901},
       number = {2},
          eid = {93},
        pages = {93},
          doi = {10.3847/1538-4357/abafbe},
archivePrefix = {arXiv},
       eprint = {2008.07469},
 primaryClass = {astro-ph.SR},
       adsurl = {https://ui.adsabs.harvard.edu/abs/2020ApJ...901...93B}
}

@ARTICLE{Catalan+2008,
       author = {{Catal{\'a}n}, S. and {Isern}, J. and {Garc{\'\i}a-Berro}, E. and {Ribas}, I.},
        title = "{The initial-final mass relationship of white dwarfs revisited: effect on the luminosity function and mass distribution}",
      journal = {\mnras},
         year = 2008,
        month = jul,
       volume = {387},
       number = {4},
        pages = {1693-1706},
          doi = {10.1111/j.1365-2966.2008.13356.x},
archivePrefix = {arXiv},
       eprint = {0804.3034},
 primaryClass = {astro-ph},
       adsurl = {https://ui.adsabs.harvard.edu/abs/2008MNRAS.387.1693C}
}

@INPROCEEDINGS{Dufour+2017,
       author = {{Dufour}, P. and {Blouin}, S. and {Coutu}, S. and {Fortin-Archambault}, M. and {Thibeault}, C. and {Bergeron}, P. and {Fontaine}, G.},
        title = "{The Montreal White Dwarf Database: A Tool for the Community}",
    booktitle = {20th European White Dwarf Workshop},
         year = 2017,
       editor = {{Tremblay}, P. -E. and {Gaensicke}, B. and {Marsh}, T.},
       series = {Astronomical Society of the Pacific Conference Series},
       volume = {509},
        month = mar,
        pages = {3},
          doi = {10.48550/arXiv.1610.00986},
archivePrefix = {arXiv},
       eprint = {1610.00986},
 primaryClass = {astro-ph.SR},
       adsurl = {https://ui.adsabs.harvard.edu/abs/2017ASPC..509....3D}
}

@ARTICLE{Nordhaus+2011,
       author = {{Nordhaus}, J. and {Wellons}, S. and {Spiegel}, D.~S. and {Metzger}, B.~D. and {Blackman}, E.~G.},
        title = "{Formation of high-field magnetic white dwarfs from common envelopes}",
      journal = {Proceedings of the National Academy of Science},
         year = 2011,
        month = feb,
       volume = {108},
       number = {8},
        pages = {3135-3140},
          doi = {10.1073/pnas.1015005108},
archivePrefix = {arXiv},
       eprint = {1010.1529},
 primaryClass = {astro-ph.SR},
       adsurl = {https://ui.adsabs.harvard.edu/abs/2011PNAS..108.3135N}
}

@ARTICLE{Garcia-Berro+2012,
       author = {{Garc{\'\i}a-Berro}, Enrique and {Lor{\'e}n-Aguilar}, Pablo and {Aznar-Sigu{\'a}n}, Gabriela and {Torres}, Santiago and {Camacho}, Judit and {Althaus}, Leandro G. and {C{\'o}rsico}, Alejandro H. and {K{\"u}lebi}, Baybars and {Isern}, Jordi},
        title = "{Double Degenerate Mergers as Progenitors of High-field Magnetic White Dwarfs}",
      journal = {ApJ},
         year = 2012,
        month = apr,
       volume = {749},
       number = {1},
          eid = {25},
        pages = {25},
          doi = {10.1088/0004-637X/749/1/25},
archivePrefix = {arXiv},
       eprint = {1202.0461},
 primaryClass = {astro-ph.SR},
       adsurl = {https://ui.adsabs.harvard.edu/abs/2012ApJ...749...25G}
}

@article{Schreiber,
author = {Schreiber, M. and Belloni, Diogo and Gänsicke, Boris and Parsons, Steven and Zorotovic, Monica},
year = {2021},
month = {07},
pages = {1-7},
title = {The origin and evolution of magnetic white dwarfs in close binary stars},
volume = {5},
journal = {Nature Astronomy},
doi = {10.1038/s41550-021-01346-8}
}

@INPROCEEDINGS{Garcia2013,
       author = {{Garc{\'\i}a-Berro}, E. and {Torres}, S. and {Lor{\'e}n-Aguilar}, P. and {Aznar-Sigu{\'a}n}, G. and {Camacho}, J. and {K{\"u}lebi}, B. and {Isern}, J. and {Althaus}, L.~G. and {C{\'o}rsico}, A.~H.},
        title = "{On the Origin of High-field Magnetic White Dwarfs}",
    booktitle = {18th European White Dwarf Workshop.},
         year = 2013,
       editor = {{Krzesi{\'n}ski}, J. and {Stachowski}, G. and {Moskalik}, P. and {Bajan}, K.},
       series = {Astronomical Society of the Pacific Conference Series},
       volume = {469},
        month = jan,
        pages = {423},
          doi = {10.48550/arXiv.1209.2622},
archivePrefix = {arXiv},
       eprint = {1209.2622},
 primaryClass = {astro-ph.SR},
       adsurl = {https://ui.adsabs.harvard.edu/abs/2013ASPC..469..423G}
}

@ARTICLE{Salmeron2007MNRAS.375..177S,
       author = {{Salmeron}, Raquel and {K{\"o}nigl}, Arieh and {Wardle}, Mark},
        title = "{Angular momentum transport in protostellar discs}",
      journal = {\mnras},
         year = 2007,
        month = feb,
       volume = {375},
       number = {1},
        pages = {177-183},
          doi = {10.1111/j.1365-2966.2006.11277.x},
archivePrefix = {arXiv},
       eprint = {astro-ph/0611359},
 primaryClass = {astro-ph},
       adsurl = {https://ui.adsabs.harvard.edu/abs/2007MNRAS.375..177S}
}

@misc{roepke2022simulationscommonenvelopeevolutionbinary,
      title={Simulations of common-envelope evolution in binary stellar systems: physical models and numerical techniques}, 
      author={Friedrich K. Roepke and Orsola De Marco},
      year={2022},
      eprint={2212.07308},
      archivePrefix={arXiv},
      primaryClass={astro-ph.SR},
      url={https://arxiv.org/abs/2212.07308}, 
}

@article{Bagnulo_2022,
	author = {Bagnulo, Stefano and Landstreet, John D.},
	doi = {10.3847/2041-8213/ac84d3},
	journal = {The Astrophysical Journal Letters},
	month = {aug},
	number = {1},
	pages = {L12},
	publisher = {The American Astronomical Society},
	title = {Multiple Channels for the Onset of Magnetism in Isolated White Dwarfs},
	url = {https://doi.org/10.3847/2041-8213/ac84d3},
	volume = {935},
	year = {2022}}

@article{Brandenburg_2005,
   title={Astrophysical magnetic fields and nonlinear dynamo theory},
   volume={417},
   ISSN={0370-1573},
   url={http://dx.doi.org/10.1016/j.physrep.2005.06.005},
   DOI={10.1016/j.physrep.2005.06.005},
   number={1–4},
   journal={Physics Reports},
   publisher={Elsevier BV},
   author={Brandenburg, Axel and Subramanian, Kandaswamy},
   year={2005},
   month=oct, pages={1–209} }

@article{Ginzburg_2022,
   title={Slow convection and fast rotation in crystallization-driven white dwarf dynamos},
   volume={514},
   ISSN={1365-2966},
   url={http://dx.doi.org/10.1093/mnras/stac1363},
   DOI={10.1093/mnras/stac1363},
   number={3},
   journal={Monthly Notices of the Royal Astronomical Society},
   publisher={Oxford University Press (OUP)},
   author={Ginzburg, Sivan and Fuller, Jim and Kawka, Adela and Caiazzo, Ilaria},
   year={2022},
   month=may, pages={4111–4119} }

@article{10.1093/mnras/stw1619,
    author = {Bhat, Pallavi and Ebrahimi, Fatima and Blackman, Eric G.},
    title = "{Large-scale dynamo action precedes turbulence in shearing box simulations of the magnetorotational instability}",
    journal = {Monthly Notices of the Royal Astronomical Society},
    volume = {462},
    number = {1},
    pages = {818-829},
    year = {2016},
    month = {07},
    issn = {0035-8711},
    doi = {10.1093/mnras/stw1619},
    url = {https://doi.org/10.1093/mnras/stw1619},
    eprint = {https://academic.oup.com/mnras/article-pdf/462/1/818/18470481/stw1619.pdf},
}

@ARTICLE{Balbus1998,
       author = {{Balbus}, Steven A. and {Hawley}, John F.},
        title = "{Instability, turbulence, and enhanced transport in accretion disks}",
      journal = {Reviews of Modern Physics},
         year = 1998,
        month = jan,
       volume = {70},
       number = {1},
        pages = {1-53},
          doi = {10.1103/RevModPhys.70.1},
       adsurl = {https://ui.adsabs.harvard.edu/abs/1998RvMP...70....1B}
}

@ARTICLE{Shakura1973,
       author = {{Shakura}, N.~I. and {Sunyaev}, R.~A.},
        title = "{Black holes in binary systems. Observational appearance.}",
      journal = {Astronomy and Astrophysics},
         year = 1973,
        month = jan,
       volume = {24},
        pages = {337-355},
       adsurl = {https://ui.adsabs.harvard.edu/abs/1973A&A....24..337S}
}

@article{bagnulo2020,
    author = {Bagnulo, Stefano and Landstreet, John D.},
    title = {Discovery of six new strongly magnetic white dwarfs in the 20 pc local population},
    DOI = "10.1051/0004-6361/202038565",
    url = "https://doi.org/10.1051/0004-6361/202038565",
    journal = {Astronomy and Astrophysics},
    year = 2020,
    volume = 643,
    pages = "A134",
}

@BOOK{KippenahanWiegert,
       author = {{Kippenhahn}, Rudolf and {Weigert}, Alfred},
        title = "{Stellar Structure and Evolution}",
         year = 1990,
       publisher = {Springer-Verlag},
       adsurl = {https://ui.adsabs.harvard.edu/abs/1990sse..book.....K}
}

@article{ Berdyugin2023,
	author = {{Berdyugin, Andrei V.} and {Piirola, Vilppu} and {Bagnulo, Stefano} and {Landstreet, John D.} and {Berdyugina, Svetlana V.}},
	title = {Discovery of magnetic fields in five DC white dwarfs},
	DOI= "10.1051/0004-6361/202245149",
	url= "https://doi.org/10.1051/0004-6361/202245149",
	journal = {A\&A},
	year = 2023,
	volume = 670,
	pages = "A2",
}

@ARTICLE{Berdyugin2022,
       author = {{Berdyugin}, Andrei V. and {Piirola}, Vilppu and {Bagnulo}, Stefano and {Landstreet}, John D. and {Berdyugina}, Svetlana V.},
        title = "{Highly sensitive search for magnetic fields in white dwarfs using broad-band circular polarimetry}",
      journal = {Astronomy and Astrophysics},
         year = 2022,
        month = jan,
       volume = {657},
          eid = {A105},
        pages = {A105},
          doi = {10.1051/0004-6361/202142173},
archivePrefix = {arXiv},
       eprint = {2111.11174},
 primaryClass = {astro-ph.SR},
       adsurl = {https://ui.adsabs.harvard.edu/abs/2022A&A...657A.105B}
}

@misc{berdyugin2024searchingmagneticfieldsfeatureless,
      title={Searching for magnetic fields in featureless white dwarfs with the DIPOL-UF polarimeter at the Nordic Optical Telescope}, 
      author={A. Berdyugin and J. D. Landstreet and S. Bagnulo and V. Piirola and S. V. Berdyugina},
      year={2024},
      eprint={2408.03614},
      archivePrefix={arXiv},
      primaryClass={astro-ph.SR},
      url={https://arxiv.org/abs/2408.03614}, 
}

@misc{ginzburg2024youngerageoldestmagnetic,
      title={Younger age for the oldest magnetic white dwarfs}, 
      author={Sivan Ginzburg},
      year={2024},
      eprint={2408.04695},
      archivePrefix={arXiv},
      primaryClass={astro-ph.SR},
      url={https://arxiv.org/abs/2408.04695}, 
}

@article{Blatman2024,
    author = {Blatman, Daniel and Ginzburg, Sivan},
    title = "{Magnetic field breakout from white dwarf crystallization dynamos}",
    journal = {Monthly Notices of the Royal Astronomical Society},
    volume = {528},
    number = {2},
    pages = {3153-3162},
    year = {2024},
    month = {01},
    issn = {0035-8711},
    doi = {10.1093/mnras/stae222},
    url = {https://doi.org/10.1093/mnras/stae222},
    eprint = {https://academic.oup.com/mnras/article-pdf/528/2/3153/56544140/stae222.pdf},
}


\end{document}